\documentstyle[11pt]{article}
\newlength{\dinwidth}                                                           
\newlength{\dinmargin}                                                          
\def\e{{\bf e}}
\def\fg{{\bf f}}
\def\h{{\bf h}}
\def\a{\alpha}
\def\b{\beta}
\def\g{\gamma}
\def\d{\delta}
\def\Ai{A_{\i}}

\def\C{{\bf C}}
\def\sbo{{\widetilde{ Obs} }}
\def\obs{{ Obs}}
\def\bi{\bibitem}
\def\ti{\tilde}
\def\tr{{\rm tr}}
\def\i{\infty}
\def\E{{\cal E}}

\def\M{{\cal M}}

\def\cO{{\cal O}}

\def\P{\Psi}
\def\x{\xi}
\def\xb{{\bar{\xi}}}
\def\zb{{\bar z}}
\def\D{\Delta}

\def\be{\begin{equation}}
\def\coset{SL(2,\R)/SO(2)}

\def\ee{\end{equation}}
\def\la{\label}
\def\c{\cite}
\def\f{\frac}
\def\Eb{\bar{\cal E}}
\def\ba{\begin{array}}
\def\ea{\end{array}}
\def\r{\ref}
\def\Ref#1{(\ref{#1})}
\def\l{w}
\def\I{{\bf 1}}

\def\o{\omega}
\def\O{\Omega}

\def\t{\gamma}

\def\Psit{\tilde{\Psi}}

\def\R{{\bf R}}
\def\cC{{\cal C}}
\def\Cx{{\cC}^{(\x)}}
\def\tz{z}
\def\Cxb{{\cC}^{(\xb)}}
\def\Psit{\tilde{\Psi}}

\def\baa{\begin{array}}
\def\eaa{\end{array}}
\def\ba{\begin{eqnarray}}
\def\ea{\end{eqnarray}}

\def\ft#1#2{{\textstyle {\frac{#1}{#2}} }}
\def\s{\sigma}

\def\bi{\bibitem}
\def\Hx{H^{(\x)}}
\def\Hxb{H^{(\xb)}}
\def\H{{\cal H}}
\def\p{\partial}
\def\0{|p \rangle }
\def\k{2k}
\def\phi{\varphi}
\def\E{{\cal E}}

\def\Eb{\bar{\cal E}}
\def\ra{\rightarrow}

\def\Aj0{A_j^{(0)}}

\def\Tj0{T_j^{(0)}}

\def\Hwj{H^{(w_j)}}

\def\si{s_{\i}}

\begin{document}

\newtheorem{Theorem}{Theorem}
\newtheorem{Lemma}{Lemma}
\newtheorem{Statement}{Statement}
\newtheorem{Definition}{Definition}
\newtheorem{Corollary}{Corollary}
\newtheorem{Assumption}{Assumption}
\vskip1.0cm
\begin{flushright} DESY 96-091 \\
                   hep-th/9605144   \end{flushright}
\vskip2.0cm
\begin{center}

\LARGE{Isomonodromic Quantization of Dimensionally Reduced Gravity}

\vskip2.0cm

\large{D.~Korotkin\footnote{On leave of absence from
Steklov Mathematical Institute, Fontanka, 27, St.Petersburg, 191011
Russia.}
 and H.~Nicolai}
\vskip1.0cm
II Institute for Theoretical Physics, Hamburg University \\
Luruper Chaussee 149, Hamburg 22761 Germany

\end{center}

\vskip3.0cm

\noindent
{\large {\bf Abstract.}} We present a detailed account of the isomonodromic 
quantization of dimensionally reduced Einstein gravity with
two commuting Killing vectors. This theory constitutes
an integrable ``midi-superspace" version of quantum gravity
with infinitely many interacting physical degrees of freedom.
The canonical treatment is based on the complete separation of
variables in the isomonodromic sectors of the model. 
The Wheeler-DeWitt and diffeomorphism constraints are
thereby reduced to the Knizhnik-Zamolodchikov equations 
for $SL(2,\R)$. The physical states are shown to live in a well 
defined Hilbert space and are manifestly invariant under the 
full diffeomorphism group. An infinite set of independent 
observables \`a la Dirac exists both at the classical 
and the quantum level. Using the discrete unitary representations
of $SL(2,\R)$, we construct explicit quantum states. However,  
satisfying the additional constraints associated with the coset 
space $\coset$ requires solutions based on the principal 
series representations, which are not yet known. We briefly discuss the  
possible implications of our results for string theory.

\newpage

\section{Introduction}
\setcounter{equation}{0}

The purpose of this article is to explain in detail the
new Hamiltonian formulation of dimensionally reduced gravity
presented in \c{KN1} and to study its exact quantization as an integrable
model of quantum gravity with infinitely many physical degrees of
freedom on the basis of the methods introduced in \c{KN2}.
Our results generalize to the more general models 
that one would obtain by dimensional reduction of certain matter 
coupled models of gravity and supergravity, but we will in this paper
deal only with the stationary axisymmetric reduction of Einstein's
theory, deferring the discussion of the Lorentzian signature
(colliding plane wave) case as well as of arbitrary gravitationally
coupled $\sigma$-models and the extension to locally supersymmetric
models to another publication. This we do mainly in order to
keep the technical complexity of the construction to
a minimum and to bring out the salient features as clearly as possible.

As is well known our general understanding of the mathematical and
conceptual problems of quantum gravity is severely hampered by the 
scarcity of ``realistic" exact solutions of the Wheeler-DeWitt 
(WDW) equation (for introductory reviews of the subject 
with many further references see \c{Hawking, QG, Ashtekar}). 
The known examples of exactly solvable models
include pure gravity \c{Witten} (see also \c{Carlip})
and supergravity \c{dWMN} in three dimensions, as well 
as certain mini-superspace models such as static spherically symmetric
gravity \c{KTh} or supersymmetric  models of the type 
considered in \c{Graham} and references therein. 
These models describe only finitely many physical 
degrees of freedom, and one may thus suspect that they are 
bound to miss essential features of quantum gravity. 
It is therefore desirable to find models with infinitely many physical 
degrees of freedom. An example of such a model is the
quantum theory of cylindrical gravitational waves \c{Kuchar,Pierri} 
corresponding to a truncation of dimensionally
reduced gravity for which the Ernst potential is real and
\Ref{ee} below can be transformed into a free wave equation.

Our main intention here is to demonstrate that models with infinitely
many {\em self-interacting} physical degrees of freedom can be treated exactly,
and that the methods which have been developed over many years 
in the context of flat space integrable systems \c{FT, Korep} 
can be transplanted to quantum gravity, yielding  a class of 
completely integrable ``midi-superspace" models (which reduces to
the Euclidean version of cylindrical gravitational waves for 
abelian groups). With regard to the 
conceptual problems of quantum gravity we shall adopt 
the pragmatic attitude that knowledge of sufficiently complicated 
exact solutions of the type constructed here
may furnish new and essential insights also with regard to the proper
interpretation of the ``wave function of the universe". Indeed,
notwithstanding the remaining technical difficulties, certain
generic problems of quantum gravity are neatly 
resolved in our model. First of all there is a well defined Hilbert space
for each isomonodromic sector and for all soliton numbers $N$. 
Although details remain to be worked out when $N$ becomes 
infinite or even continuous, it is clear at least in principle 
how to construct the full Hilbert space as an inductive limit. 
Secondly, the fact that the WDW equation and diffeomorphism constraint
express the invariance of the quantum state w.r.t. the  full
set of $2d$ coordinate transformations is completely manifest;
the scalar product which naturally exists in the 
isomonodromic subspaces is positive definite for unitary
representations and respects the full diffeomorphism 
invariance upon restriction to the subspace of physical states.
Finally, while the construction of observables \`a la Dirac 
remains a largely unsolved problem of canonical gravity 
in general \c{QG}, it turns out that our model admits an  
infinite number of independent ones, namely the monodromies 
associated with the singularities of the logarithmic derivative 
of $\Psi$ in the spectral parameter plane: these are the conserved
``non-local charges" of matter coupled quantum gravity. 
The correlators of observables
-- the only meaningful expectation values in quantum gravity -- 
are well defined and can be computed at least in principle. 

Our treatment of axisymmetric stationary
quantum gravity relies in an essential way on the novel canonical
formulation of the Ernst equation proposed in \c{KN1} and is based on
the complete separation of the equations of motion and the use
of the logarithmic derivative of the related $\Psi$-function with 
respect to the spectral parameter as the fundamental canonical variable. 
In \c{KN1} we have proved that, in the classical theory, 
the conformal factor is essentially the $\tau$-function 
associated with the Ernst equation. Furthermore,
as shown in \c{KN2}, the WDW equation for this class of models  
can be reduced to the Knizhnik-Zamolodchikov (KZ) equations \c{KZ}. 
While only the gravitationally coupled principal 
chiral $SU(2)$ model was analyzed from this point of view in \c{KN2}
we will here extend these considerations to coset spaces and
non-compact groups. Completely explicit exact solutions 
of the WDW equation based on the discrete unitary representations 
of $SU(2)$ and $SL(2,\R)$, respectively, are thereby obtained.
Unfortunately, we are not able so far to
solve the additional coset constraints for the non-compact 
space $\coset$ with the discrete unitary representations of $SL(2,\R)$;
rather, it appears that one will have to make use 
of the principal continuous series representations of $SL(2,\R)$,
for which no solutions of the KZ equations are known so far. 
This is the major open problem left by the present work.

We also hope that these results may eventually
enable us to address some other unsolved problems of 
current research in the framework of exactly solvable models.
Our methods can be generalized without difficulty of principle 
to matter coupled gravity and supergravity because matter 
and gravity are unified in the group theoretical construction 
(for instance, to quantize Maxwell Einstein gravity one simply 
would have to replace the coset space $SL(2,\R)/SO(2)$ by $SU(2,1)/U(2)$).
Since the exact WDW functionals are built on classical
solutions, we can in principle obtain solutions of the WDW equation 
which in a very precise sense are ``close'' to a given classical solution 
of Einstein's equations and study their $\hbar\ra 0$ limits.
Understanding the semiclassical limit is also a necessary prerequisite for
explaining the UV divergences that would appear in a conventional 
perturbative treatment and that are invisible in the isomonodromic 
sectors which are ``far away" from the perturbative regime.
An intriguing aspect of our work is
the possible relevance of quantum groups suggested by the link with
the KZ equations; it appears that the notion
of quantum space-time \c{Wess} may emerge quite naturally 
in the present framework. Our results might also
shed some light on the information loss
paradox for Hawking radiation: in view of persisting 
disagreements \c{Banks}, one cannot help feeling that the Gordian 
knot can only be cut by finding an exactly solvable model for it.

In section 2 we briefly recall the origin of our model 
as Kaluza-Klein-like dimensional reduction of $4d$ 
Einstein's equations with two commuting Killing vectors. 
Section 3 gives a classical treatment of the model in the framework of the
inverse scattering method. In the isomonodromic sector of the model we
rewrite the equations of motion as a system of Schlesinger-like
deformation equations by introducing new canonical variables, and present the
two-time Hamiltonian formulation suitable for quantization.
In Section 4 we explicitly quantize this Poisson structure. The 
link between the WDW equations and the KZ equations for $SU(2)$
and $SL(2,\R)$ is established. For the standard unitary 
representations of $SU(2)$ and the unitary discrete
series representations of $SL(2,\R)$ this allows us to write down
the exact WDW functional in terms of the integral representations 
for solutions of the KZ equations and to define the  
quantum $\tau$-function. We carefully explain the constraints
that must be satisfied when one passes from the group 
space $SL(2,\R)$ to the coset space $\coset$ which put in 
evidence the need for continuous (principal series)
unitary representations of $SL(2,\R)$. We also discuss possible ways 
to construct physically important solutions in the classical limit,
such as the Kerr-NUT solution. Since many of our results have 
a decidedly stringy flavor, we elaborate a little on the 
hints pointing towards the existence of a new kind of dual model
in a separate section 5. In the appendix we summarize the results about 
unitary representations of $SL(2,{\bf R})$ needed in the main text. 

\section{Axisymmetric stationary gravity as a nonlinear $\sigma$-model}
\setcounter{equation}{0}

We will proceed from the standard metric of a stationary
axisymmetric $4d$ space-time
\be
ds^2 = f^{-1}[e^{2k} (dx^2+d\rho^2)+\rho^2d\phi^2]   
-f (dt + F d\phi)^2    
\la{metric}\ee
Here $(x,\rho)$ are canonical Weyl coordinates with $\rho \geq 0$
and $x\in \R$, and $t$ and $\phi$ are
time and angular coordinates, respectively. By assumption, the
functions appearing in \Ref{metric} depend only on $(x,\rho)$,
and the coordinates $t$ and $\phi$ thus play no role in the remainder.
The metric coefficients appearing in \Ref{metric} are commonly 
expressed in terms of the so-called Ernst potential $\E$
\c{Ernst, Exact} as follows:
\be
k_{\x}=2i\rho\f{\E_{\x}\Eb_{\x}}{(\E+\Eb)^2}\;\;\;\;\;\;\;\;
F_{\x}=2\rho\f{(\E-\Eb)_{\x}}{(\E+\Eb)^2}\;\;\;\;\;\;\;
f={\rm Re} \, \E
\la{dual}\ee
where $\x:=x+i\rho\, ,\, \xb:=x-i\rho$, and subscripts
stand for partial derivatives throughout this paper. 
Next we define the symmetric matrix
\be
g=\f{1}{\E+\Eb}
    \pmatrix{ 2 & i(\E-\Eb)\cr
            i(\E-\Eb) & 2\E\Eb  \cr}
\la{g}\ee
which can be viewed as an element of the coset space $SL(2,\R)/SO(2)$.
Einstein's equations then imply the Ernst equation \c{Exact}
\be
\big((\x-\xb)g_{\x}g^{-1}\big)_{\xb} + 
\big((\x-\xb)g_{\xb}g^{-1}\big)_{\x} =0
\la{ee}\ee
which in this form closely resembles the equation that one would
obtain for a principal chiral model. Further, they
give rise to the following first order equations 
for the conformal factor $h\equiv e^{2k}$
\be
2 k_\x =(\log h)_{\x}=\f{\x-\xb}{4}\tr(g_{\x} g^{-1})^2\;\;\;\;\;\;
2k_\xb =(\log h)_{\xb}=\f{\xb-\x}{4}\tr(g_{\xb}g^{-1})^2
\la{h1}\ee
We will find it convenient to reexpress these equations by means of
the one-form $\o$ defined by 
\be
\o=\f{\x-\xb}{4}\tr (g_{\x} g^{-1})^2 d\x +
\f{\xb-\x}{4}\tr (g_{\xb} g^{-1})^2 d\xb
\la{om}\ee
Then \Ref{h1} simply becomes
\be
  dk = \ft12 \o
\la{h}
\ee
Both \Ref{h1} and \Ref{h} are consistently defined 
because \Ref{ee} implies $d\o =0$. We note already here
that the equations \Ref{h1} will turn into (linear combinations of) the 
WDW equation and the diffeomorphism constraint upon quantization.

The above equations of motion, which were obtained by dimensional
reduction on Einstein's equations, can be alternatively derived directly 
from a $SL(2,\R)/SO(2)$ coset space $\sigma$-model in two space-time
dimensions coupled to $2d$ gravity and a dilaton field $\rho$ \c{KraNeu}. 
If the Euclidean worldsheet is locally
parametrized by the complex coordinates $(z,\zb)$,
the metric has the following form in the conformal gauge:
\be
ds^2=h(z,\zb)dz d\zb \;\;\;\;\;\;\;\;\;\;h\equiv e^{\k}
\la{metr}\ee
A suitable Lagrangian is 
\begin{equation}
{\cal L} =\rho \Big( h R + {\rm tr} (g_{z} g^{-1}
g_{\zb}g^{-1}) \Big)
\la{L}\end{equation}
where $R$ is the Gaussian curvature of the worldsheet, i.e.
$$ R=(\log h)_{z\zb}/h  , $$
and the trace in \Ref{L} is appropriately normalized. As is well known
from string theory, the first order equations \Ref{h1} can be obtained
from \Ref{L} by variation with respect to the off-diagonal elements
of the worldsheet metric, so the conformal gauge condition must be
temporarily relaxed. It is then obvious that these equations are
completely analogous to the Virasoro conditions of string theory.

In the Lagrangian \Ref{L} the dilaton $\rho$ appears as an independent
field, not as a coordinate. To establish the relation
with the previous formulation for the metric \Ref{metric} 
in terms of Weyl canonical coordinates we note that the equation
of motion for $\rho$ following from \Ref{L} 
\be
\rho_{z\zb} = 0      \la{dilequ}
\ee
is solved by 
\be
\rho (z,\zb) = {\rm Im} \, \x( z)
\la{gaugefix}\ee
where $\x(z)$
is a (locally) holomorphic function, and that the conformal gauge \Ref{metr} 
is left intact by holomorphic reparametrizations of the complex coordinate
$z$ (for which the metric remains diagonal and 
$h(z,\zb)$ is simply multiplied by a factor). 
Hence we can further specialize the gauge by identifying 
$\rho$ with one of the worldsheet coordinates as in \Ref{metric}
(global aspects of this change of variables were discussed in \c{KN3}
but will not concern us here). From the $2d$ point of view, the
gravitational field $h$ and the matter field $g$ are coupled through the
dilaton $\rho$; furthermore, the obvious solution $\rho= const$ of
(\r{dilequ}) would imply $g_{\x}=g_{\xb}=h_{\x}= h_{\xb}=0$,
i.e. the trivial solution for the matter fields, in which case the
gravitational sector would become purely topological (strictly speaking,
this argument relies on the positive definiteness of the Cartan Killing
metric on the coset $SL(2,\R)/SO(2)$). Therefore the model possesses no
non-trivial flat space limit since the matter fields
act as sources for $2d$ gravity and thus distort the
two-dimensional background geometry.

Although we do not know  whether a ``Wick rotation" 
can be rigorously justified in the context of (quantum) 
gravity (see, however, \c{Thiemann} for a recent discussion), 
we will occasionally take the liberty to refer to
$\rho$ and $x$ as ``time" and ``space" coordinates, respectively,
especially in connection with the canonical treatment.

\section{Classical treatment}
\setcounter{equation}{0}

\subsection{Linear system and isomonodromy}
The results of this and the following section 
apply  to
solutions $g(\x,\xb)$ of (\r{ee}) belonging to the 
complex general linear group $GL(n,\C)$. 
The extra conditions needed to ensure that $g(\x,\xb)$ is an 
element of the coset space $G/H$ and that 
$h(\x,\xb) \in\R$ (as required by dimensionally reduced gravity) 
will be presented in detail in section 3.3.
 
Equation (\r{ee}) is the compatibility condition of the following
linear system  \c{M,BZ}:
\be
\f{d\P}{d \x}
=\f{g_{\x} g^{-1}}{1-\t}\P\;\;\;\;\;\;\;\; ,
\;\;\;\;\;\;\;\;\;
\f{d\P}{d \xb}
=\f{g_{\xb} g^{-1}}{1+\t}\P
\la{ls}\ee
where $\P(\t;\x,\xb)$ is a two-by-two matrix 
from which the metric (\r{metric}) can be reconstructed.
The function $\t(\x,\xb)$ is a ``variable spectral parameter"
subject to the following (compatible) first order equations
\be
\t_{\x}=\f{\t}{\x-\xb}\f{1+\t}{1-\t} \;\;\;\; , \;\;\;\;
\t_{\xb}=\f{\t}{\xb-\x}\f{1-\t}{1+\t} .
\la{pe}\ee
They are solved by
\be
\t_\pm (\l ; \x,\xb)
  =\f{2}{\x-\xb}\left\{\l-\f{\x+\xb}{2} \pm\sqrt{(\l-\x)(\l-\xb)}\right\}
  =\f{1}{\t_\mp (\l; \x,\xb)}
\la{gamma}\ee
with $\l\in {\bf C}$ a constant of integration, which can be
regarded as the ``hidden" constant spectral parameter. In the
sequel we will usually suppress the index $\pm$ and simply
write $\t(w;\x,\xb) \equiv \t_+(w;\x,\xb)$. The relation
\Ref{gamma} can be inverted to give 
\be 
w = \ft14 (\x -\xb) \Big( \t + \f{1}{\t} \Big) + \ft12 (\x + \xb)  \la{w}
\ee
which shows that, in our conventions, $\t$ is purely
imaginary for real $w$. For the
linear system (\r{ls}), we can use either $\t$ or $\l$; when
$\t$ is expressed as a function of $\l$ according to (\r{gamma}), the
linear system (\r{ls}) lives on the two-sheeted Riemann surface
of the function $\sqrt{(\l-\x)(\l-\xb)}$. Furthermore,
\be
\f{d}{d\x}\equiv\f{\partial}{\partial \x}+
\f{\t}{\x-\xb}\f{1+\t}{1-\t} \f{\partial}{\partial \t}\;\;\;\;\;\;\;\;\;
\f{d}{d\xb}\equiv\f{\partial}{\partial \xb}+
\f{\t}{\xb-\x}\f{1-\t}{1+\t} \f{\partial}{\partial \t}.
\la{cd}
\ee   
{}From (\r{ls}) we immediately obtain (see also \c{NK})
\begin{Lemma}
The following relations hold
\be
g_{\x}g^{-1}=\f{2}{\x-\xb}\P_{\t}\P^{-1}\Big|_{\t=1} \;\;\;\;\;\;\;\;
g_{\xb}g^{-1}=\f{2}{\x-\xb}\P_{\t}\P^{-1}\big|_{\t=-1}
\la{nk}\ee
where the subscript $\t$ denotes partial
differentiation with respect to $\t$.
\end{Lemma}

Next we consider the behavior of 
$(d\P/d\x)\P^{-1}$ and $(d\P/d\xb)\P^{-1}$ in the complex $\t$-plane.
The following theorem is a standard consequence of the 
formulation of classical integrable systems as a Riemann-Hilbert problem.

\begin{Theorem}
Let the two-by-two matrix $\P(\t;\x,\xb)$
be subject to the following conditions:
\begin{enumerate}

\item  As a function of $\t$ the matrix
$\P$ is holomorphic and invertible everywhere 
on some cover of the complex $\t$-plane with the exception
of the points mentioned below.
\item $\P$ has regular singularities at the branch points
$\t_j(\x,\xb):= \t(w_j;\x,\xb)$ for $j=1,...,N$ with constants
$\l_j\in \C$, in the vicinity of which it behaves as
\be
\P(\t;\x,\xb)=
G_j(\x,\xb)\P_j(\t;\x,\xb)(\t-\t_j)^{T_j} C_j\;\;\;\;{\rm as}\;\;
\;\;\t\sim \t_j
\la{rs}\ee
where, for $\t \sim \t_j$, $\P_j(\t;\x,\xb)={\bf 1} + O(\t-\t_j)$ is
holomorphic and invertible. The matrices $C_j$ and
$T_j$ are constant and invertible, and constant diagonal,
respectively, while the $(\x,\xb)$-dependent matrices $G_j$
are assumed to be invertible.

\item Across certain ``movable" contours $\{L_j\}$, which connect
the singular points $\t_j$ to some arbitrarily chosen but fixed
and non-singular base point $\t_0 \equiv \t(w_0;\x,\xb)$
and whose dependence on $(\x,\xb)$ is determined by \Ref{gamma},
the boundary values of $\P^-(\t)$ and $\P^+(\t)$ are related by
\be
\P^+(\t;\x,\xb)= \P^-(\t;\x,\xb) M_j(\l) \;\;,\;\;\;\; \t\in L_j
\la{RH}\ee
where the invertible matrices $M_j$ depend only on
the constant spectral parameter $\l$.

\item The following asymptotic conditions hold:
\be
\P(\t;\x,\xb)= g_{\i}  +O\bigg(\f{1}{\t} \bigg)
%G_\i ({\bf 1} +O\big(\t^{-1}\big)) w^{T_\i} C_\i  
\;\;\;{\rm as}\;\;\;\t\sim \i
\la{i}\ee
\be  
\P(\t;\x,\xb)= g(\x,\xb)  +O\big(\t\big)
\;\;\;{\rm as}\;\;\;\t\sim 0
\la{0}\ee
where the matrix $g_{\i}$ is constant and invertible and
the matrix $g(\x,\xb)$ is invertible. 
\end{enumerate}
Then $\P$ obeys the linear system (\r{ls}) and  
$g(\x,\xb)\in GL(2,\C)$ solves (\r{ee}). 
\la{main}
\end{Theorem}
{\it Proof.} Conditions 1.-4. imply the analyticity of
$\P_\x \P^{-1}$ and $\P_{\xb}\P^{-1} $ in $\t$ away from $\t=\pm 1$. 
The poles at $\t = \pm 1$ on the r.h.s. of \Ref{ls}
are produced solely by differentiation 
of the spectral parameter $\t$ with respect to $\x$ and $\xb$,
with residues fixed by \Ref{0}. Since condition \Ref{i}
provides the normalization at $\t =\i$, the r.h.s. of \Ref{ls} is
completely determined.
$\Box$

\begin{Definition}
A solution $g(\x,\xb)$ of (\r{ee}) is called isomonodromic if the 
associated conjugation matrices $M_j$ are independent of $w$
(i.e. do not vary along $L_j$). 
\end{Definition}
Following \c{Jimbo} we will refer to the set
$\{\l_j, T_j, C_j , L_j , M_j (w)\}$ as the set of monodromy data 
of the function $\P(\t)$ and the associated solution $g(\x,\xb)$ of \Ref{ee}.

The logarithmic derivative (``spectral parameter current")
\be
A(\t ;\x,\xb)\equiv \f{\p \P}{\p \t}\P^{-1} \la{A(t)}  \ee
will play a key role in the sequel. In general,
$A(\t)$ is not single-valued as a function of
$\t$ for non-constant matrices $M_j$. It is only for 
isomonodromic solutions that the singularities of $A(\t)$
in the $\t$-plane are simple poles at $\t_j$. In this case 
$A(\t)$ is a meromorphic function, and we have
\be
A(\t) = \sum_{j=1}^N  \f{A_{j}}{\t-\t_j}
\la{de}\ee
(if the summation range is not indicated explicitly, sums are understood
to be taken over $j=1,\dots ,N$). The residues at the points $\t=\t_j$ 
are easily computed from \Ref{rs}
\be
A_j (\x,\xb) = G_j T_j G_j^{-1}
\la{Aj}\ee
The eigenvalues of $A_j$ determine
the ramification number of $\Psi$ at $\t_j$; if they are all rational,
the number of sheets of the associated Riemann surface 
is finite, otherwise infinite. The sum of the residue matrices 
\be
A_\i :=  \lim_{\t \rightarrow \i} \big( \t A(\t) \big)
      =  \sum_{j=1}^N A_j
\la{sum}\ee
governs the asymptotical behavior of $\Psi(\t)$ at infinity.
If $\Psi (\t)$ is regular at $\t = \i$ as in \Ref{i}, we have
\be \Ai = 0  \la{regularity}  \ee
Then we can choose $\t_0 = \i$ as the base point in Thm.~\r{main}.
We shall assume \Ref{regularity} to hold throughout most of this paper. 
Inserting (\r{de}) into (\r{nk}) we obtain
\be
g_{\x} g^{-1}=\f{2}{\x-\xb}\sum_{j}\f{A_j}{1-\t_j}\;\;\;\;, \;\;\;\;
g_{\xb} g^{-1}=\f{2}{\xb-\x}\sum_{j}\f{A_j}{1+\t_j}
\la{cur}\ee
This formula shows how to reconstruct the Ernst potential, and
hence the space-time metric, once the residues $A_j$ are known.

For isomonodromic solutions, the matrices $M_j (w)\equiv M_j$ are
independent of $w$ and are called
monodromy matrices of the connection $A(\t)d\t$; we have 
\be
M_j  = C_j^{-1}e^{2\pi i T_j} C_j ; 
\la{mon}\ee

By definition, they obey
\be \f{d M_j}{d\x}=\f{d M_j}{d\xb}=0 \la{monconstant} \ee
and are thus constants of motion. To express them as
as path-ordered integrals we choose the same base point $\t_0$
as in Thm.~\r{main}; then
\be
M_j := {\cal P} \exp\oint_{l_j} \Psi^{-1} \Psi_{\t} d\t
\la{monodromy} \ee
where the contour $l_j$ starts at $\t_0$, encircles
the point $\t_j(\x,\xb)$ and returns to $\t_0$; of course,
$M_j$ does not depend on the choice of $\t_0$ as is already obvious
from \Ref{mon}. For $\t_0 = \i$ we get
\be
M_j = g_\i^{-1} \bigg({\cal P}\exp\oint_{l_j}A(\t)d\t \bigg)g_\i
\la{monodromy1} \ee
The eigenvalues of the matrices $\log M_j$ and $2\pi i A_j$ are the same;
however, the explicit relation between them is highly non-local.
The ``monodromy at infinity"
\be
\prod_{j=1}^N M_j = M_\i = \exp \big( 2\pi i \Ai \big)
\la{monprod}  \ee
is equal to ${\bf 1}$ as a corollary of the assumed regularity
of $\Psi(\t)$ at $\t = \i$. 

An obvious question at this point concerns the status of the 
isomonodromic solutions among all solutions of (\r{ee}).
At first sight \Ref{de} looks like a strong constraint
on the possible solutions, but in fact it is not.
Apart from the assumed analyticity of $\P$, the only true
assumption that goes into \Ref{de} is the absence of essential
singularities of $\P$ as a function of $\t$
(this assumption must, however, be relaxed for
locally supersymmetric theories where higher order ``rigid" poles at 
$\t =\pm 1$ appear in the linear system \c{Nic1}); then \Ref{de} is
just the analog of the well known statement that the logarithmic 
derivative of an analytic function can be represented as a sum
over its poles.
In fact, almost all known exact solutions of (\r{ee}) are isomonodromic. 
So, for multisoliton solutions of Einstein's equations \c{BZ}, all of
the matrices $M_j$ are proportional to the unit matrix such that
all eigenvalues of $A_j$ are half integer, and for the finite gap
(algebro-geometric) solutions constructed in \c{KM,KOR} the matrices
$M_j$ are either anti-diagonal (i.e. with zeros on the diagonal) 
or again proportional to $\I$.
For solutions expressible in terms of Painlev\'{e} transcendents
\c{Xant} the matrices $M_j$ are triangular. 
The only examples of solutions which are not strictly isomonodromic 
in the sense that the sum in \Ref{de} is infinite 
and that there is an accumulation point of the poles at infinity
are the periodic analogs of the axisymmetric static solutions
constructed in \c{Myers,KN3}. Evidently, such solutions may
be obtained from the $N$-soliton solutions by a limiting procedure.
Indeed, we can at least in 
principle approximate an arbitrary solution of (\r{ee}) 
by isomonodromic ones (perhaps including higher order poles)
if we approximate a given non-constant function
$M_j(\l)$ by step functions. If this procedure could be justified
rigorously we could claim that our present treatment also covers
the general case where the sum in \Ref{de} is replaced by an integral.
However, in spite of some work in this direction for other 
integrable systems \c{Its}, the results obtained so far
are still inconclusive. Rigorously speaking, we are thus considering
a truncation of the total phase space by assuming \Ref{de}, and
the question of whether we can get the full phase space as a union
(in a suitable completion) of
its isomonodromic subsectors remains open for the moment.
We conjecture that a complete treatment of the model will allow 
us to rigorously justify the restriction to the isomonodromic sectors,
which is introduced in \Ref{de} ``by hand". 
Naturally this question also bears upon the precise definition
of the Hilbert space of the quantum theory.

The analyticity properties of $\Psi(\t)$ have also been discussed
extensively in \c{BM}. To understand the difference between the
approaches of \c{BZ} and \c{BM} we note that, 
together with a given $\Psi(\t)$, the function $\Psi (\t) S(w)$ 
also solves the linear system \Ref{ls} for any non-degenerate
matrix $S(w)$. While the multiplication by such a matrix does
not affect the $(\x,\xb)$-dependence, it does alter the
analyticity properties of $\Psi$ as a function of $\t$ (and hence
the prescription for extracting the solution $g(\x,\xb)$ from it).
As a consequence, the expansion \Ref{de} is also modified
because $\p_\t S = \p_w S \, \p_\t w \neq  0$. In \c{BM}
this residual freedom in the choice of $\Psi$ is completely eliminated 
by demanding $\Psi(\t)$ to be holomorphic a certain neighborhood
of the origin whose boundary always contains the points
$\t = \pm 1$ (e.g. the unit disk in the $\t$-plane),
whereas no such restriction on the location of the poles
in the $\t$-plane is made in \c{BZ}. The precise relation between
these different ``pictures" will be further clarified in section 3.3
when we discuss the coset constraints.

\subsection{Deformation equations and $\tau$-function.}
Substituting (\r{cur}) into (\r{ls}) and demanding
compatibility between (\r{ls}) and (\r{de}) we get \c{KN1}
\be
\f{\partial A_j}{\partial\x}=
        \f{2}{\x-\xb}\sum_{k\neq j}\f{[A_k,\;A_j]}{(1-\t_k)(1-\t_j)}
\;\;\;\; , \;\;\;\;
\f{\partial A_j}{\partial\xb}=
\f{2}{\xb-\x}\sum_{k\neq j}\f{[A_k,\;A_j]}{(1+\t_k)(1+\t_j)}
\la{1}\ee
We repeat that these deformation equations as well as the 
definition of the $\tau$-function to be presented below 
are  valid generally
for the groups $GL(n,\C)$.
The equations \Ref{1} may also be represented in ``Lax form", viz.
\be
\frac{\partial A_j}{\partial \x} =
   \Big[ U|_{\t=\t_j},A_j\Big]  \;\;\;\;\;\; ,  \;\;\;\;\;\;\;
\frac{\partial A_j}{\partial \xb} =
 \Big[ V|_{\t=\t_j},A_j\Big]  ,
\la{1a}\ee
where
$$ U=\f{g_{\x} g^{-1}}{1-\t}\;\;\;\;\;\;\;
V=\f{g_{\xb} g^{-1}}{1+\t} $$
This form of (\r{1}) is ``gauge-covariant" with respect
to the transformation
\be
\ti{\P}=\O(\x,\xb)\P,
\la{gau}\ee
in the sense that the transformed function $\ti \P$ 
satisfies the linear system
$d\ti{\P}/d\x = \ti{U}\ti{\P} \; , \;
d\ti{\P}/d\xb = \ti{V}\ti{\P}$ with
\be
\ti{U}=\O_{\x}\O^{-1}+\O U\O^{-1} \;\;\;\;\;\; , \;\;\;\;\;\;
\ti{V}=\O_{\xb}\O^{-1}+\O V\O ^{-1}
\la{UV1}\ee
Clearly, the matrix functions $A_j$ transform as
$A_j\rightarrow \ti{A}_j=\O A_j\O^{-1}$ under (\r{gau}).
The transformed matrices $\ti{A}_j$ then obey the same linear
system (\r{1a}) with the pair $(U,V)$ replaced by
$(\ti{U},\ti{V})$.

We are now in a position to formulate 
\begin{Theorem}
Let $\{\l_j\in{\bf C}\; ; \;j=1,...,N\}$ be an arbitrary set of complex
constants and $A_j = A_j(\x,\xb)\in gl(2,\C)$. Then 
\begin{enumerate}
\item   The two matrix differential equations (\r{1}) are compatible if 
   $\t_j=\t(\l_j;\x,\xb)$.
\item The linear system (\r{cur}) for $g(\x,\xb)$, where $\{A_j(\x,\xb)\}$
is an arbitrary solution of (\r{1}), is also compatible, and its
solution $g(\x,\xb)\in GL(2,\C)$ satisfies equation (\r{ee}).
\la{commut}
\end{enumerate}
\end{Theorem}

The compatibility of eqs.\Ref{1} can be checked by a
straighforward computation. Combining \Ref{cur} and \Ref{1} 
we recover the (complexified) Ernst equation \Ref{ee}.
$\Box$.

Remarkably, the dependence of the (complexified)
Ernst equation and its associated linear system on the
variables $\x$ and $\xb$ has been completely decoupled by Thm.~\r{commut}.
Therefore the problem of solving Einstein's equations in
this reduction has been reduced to integrating two {\it ordinary}
matrix differential equations, which are automatically
compatible unlike the original linear system (\r{ls}). All information
about the degrees of freedom is thereby encoded into the
``initial values", i.e. the set of matrices
$A_j^{(0)} \equiv A_j(\x^{(0)}, \xb^{(0)})$, where $(\x^{(0)},\xb^{(0)})$
is an arbitrarily chosen base point, and the value at any other
point can be consistently obtained by integration along any curve
connecting it to the base point\footnote{Although we are dealing
with an elliptic rather than a hyperbolic partial differential
equation, there is no paradox here because the ``initial values"
also determine the behavior at infinity via the matrix 
$\Ai \equiv \Ai^{(0)} = -\sum_j A_j^{(0)}$ (which is a constant of motion 
by Lemma 2 below) as appropriate for an elliptic boundary value problem.}.
The matrices $A_j^{(0)}$ are also 
the appropriate phase space variables for the matter sector, 
as we will see in section 3.4. Accordingly, we will
regard the functions $A_j(\x,\xb)$ rather than $\P(\t;\x,\xb)$ as
the basic quantities from now on, and relate the system
(\r{1}) directly to the (complexified) Ernst equation (\r{ee}).
With this parametrization of the phase space the isomonodromic
subsectors can be treated separately, as they
are stable with respect to the evolution equations for arbitrary
choices of the ``soliton number" $N$ and the points $w_j$.

As an immediate consequence of \Ref{1} we can construct 
integrals of motion, confirming the statements
after \Ref{monconstant}.
\begin{Lemma}
Let $\{A_j\}$ be an arbitrary solution of the system (\r{1}).
Then the variables $\Ai\equiv\sum_j A_j$, ${\rm tr}A_j$
and ${\rm tr} A_j^2$
(and thus all eigenvalues of the two-by-two matrices $A_j$)
are $(\x,\xb)$-independent, hence constants of motion.
\la{intmot}
\end{Lemma}
Notice that the lemma is in accord with
the original definition of $A_j$ in (\r{Aj}), but more general
since we now consider {\it arbitrary} solutions of (\r{1}).
As a corollary, we conclude that the sum
\be \sum_{j<k}\tr (A_j A_k) = \ft12 \tr  \Ai^2
   - \ft12 \sum_j \tr A_j^2 \la{trAjAk}  \ee
is likewise $(\x,\xb)$-independent. The above constants of motion
will give rise to observables in the canonical framework.
 
To each solution $\{A_j\}$ of (\r{1})
we can associate the following closed one-form \c{Jimbo}:
\be
\o_0(\x,\xb)= \sum_{j <  k} \tr (A_j A_k)d\log
(\t_j-\t_k)
\la{om0}\ee
where the exterior derivative $d$ is to be taken with respect to the
deformation parameters $(\x,\xb)$. The closure condition
$d\o_0 =0$ may be directly verified by use of (\r{1}) and (\r{pe}).
Following the general prescription given in \c{Jimbo}, we have

\begin{Definition}
The function $\tau(\x,\xb)$ defined by
\be
d\log\tau =\o_0
\la{tau}\ee
is called the $\tau$-function of the Ernst equation.
\end{Definition}

We will now show that the $\tau$-function has a very definite
physical meaning in our context: up to an explicit factor, it
is just the conformal factor $h\equiv e^{2k}$ !
To establish this result, we first substitute
(\r{cur}) into (\r{om}); then using (\r{pe}) and (\r{om0}) we obtain
\be
 \o=\o_0+\f{1}{\x-\xb}\sum_{j}\tr A_j^2
\left\{\f{d\x}{(1-\t_j)^2} -\f{d\xb}{(1+\t_j)^2}\right\}+
\sum_{j<k}\tr (A_j A_k)d\log (\x-\xb)
\la{f1}\ee

By \Ref{trAjAk} all extra terms on the
r.h.s. of (\r{f1}) may be explicitly integrated.
Using (\r{pe}),(\r{f1}) and the relation
$$\f{2}{\x-\xb}\left\{\left(\f{1}{(1-\t_j)^2}-\f{1}{2}\right)d\x-
\left(\f{1}{(1+\t_j)^2}-\f{1}{2}\right)d\xb\right\}=
  d\;\bigg( {\rm log}\f{\t_j^2}{(\x-\xb)(1-\t_j^2)}\bigg)=
  d\;\Big( {\rm log}\f{\p\t_j}{\p\l_j} \Big) $$
we arrive at 

\begin{Theorem}

The conformal factor $h$ defined by \Ref{om} and 
(\r{h}), and the $\tau$-function defined by (\r{tau})
are related by
\be
h\big(\x,\xb ; \{ \l_j \} \big) =
C(\x-\xb)^{\f{1}{2} \tr \Ai^2  }    \prod_j
\left\{\f{\partial \t_j}{\partial\l_j}\right\}^{\f{1}{2}\tr A_j^2}
\tau \big(\x,\xb ; \{ \l_j \}\big)
\la{link3}\ee  
where $C$ is an integration constant.
\la{confac}
\end{Theorem}
Observe that the notation in
\Ref{link3} can be further unified by writing (see \Ref{w})
$$ \f{4}{ \x - \xb } = \f{\p \t}{\p w} \bigg|_{\t =\i}. $$
For $\Ai = 0$, the first factor in \Ref{link3} can be dropped.
Also, for the group $GL(2,\C)$, the function $h(\x,\xb)$ 
is complex in general. Apart from its structural content 
this theorem is more general than previous results, 
where the conformal factor was computed only for 
multisoliton solutions \c{BZ}. It would be interesting
to find out how it is related to the ``cocycle formula" of \c{BM}.

The above formulas simplify considerably for abelian groups --
corresponding to the Euclidean version of cylindrical 
waves for $G= GL(1,\R) = \R^+$
or the so-called Gowdy models for $G=U(1)$ -- or if the
the residue functions $A_j$ are in the Cartan subalgebra of the
non-abelian group under consideration (which for $G=SL(2,\R)$
would give the multi-Schwarzschild solutions). Namely, 
if the commutators on the r.h.s. side of \Ref{1} vanish 
all $A_j$ are constant. One can then immediately write down
the solution of the linear system which reads
\be 
\Psi(\t;\x,\xb) = F(\x,\xb)
\prod_{j=1}^N \big(\t - \t_j \big)^{A_j}   \la{abelian1} \ee
The undetermined function $F(\x,\xb)$ is fixed by imposing the
proper $\t$-dependence in \Ref{ls} and given by
\be
F(\x,\xb) = \big( \x-\xb \big)^{\sum A_j}
\la{abelian2} \ee
(thus $F\equiv 1$ for $\Ai =0$). From this we can immediately 
read off the solution of \Ref{ee} with $\phi \equiv \log g$
\be 
\phi(\x,\xb) = 
  \sum_{j=1}^N A_j \log \Big( (\x-\xb) \t(w_j;\x,\xb) \Big)  
\la{abelian3} \ee
\Ref{abelian3} bears some similarity to the mode expansion of a string
target space coordinate in terms of oscillators. However, it is better 
to think of \Ref{abelian3} as an expansion in terms of coherent states.

\subsection{Reality conditions and coset constraints} 
The solutions $g(\x,\xb)$ obtained from the deformation equations \Ref{1} 
in general are neither symmetric nor even in $SL(2,\R)$. 
We must therefore impose extra conditions in order to ensure that
$g(\x,\xb) \in SL(2,\R)/SO(2)$, i.e. that $g$ is real and symmetric, 
and that the conformal factor is real.
For this purpose we need both a reality condition on the 
moving poles $\t_j$ (and thus on the parameters $w_j$) as well as 
certain extra constraints on $\P$, the monodromy matrices or the 
$\tau$-function. These conditions may be formulated in several 
equivalent ways which we shall now present. 

Quite generally it is clear that for $g$ to be in a given Lie group 
the residues $A_j$ in \Ref{de} must be elements of
the associated (complexified) Lie algebra;
consequently, for simple Lie groups, we have ${\rm tr} A_j =0$. 
Depending on which real form one is dealing with, the parameters $w_j$ 
and the matrices $A_j$ are subject to certain reality conditions.
In the case at hand, these  
follow from the simple requirement that the two expressions in 
\Ref{cur} are complex conjugate to one another for real $g$.
It is important here that only the sums on the r.h.s. of \Ref{cur} 
are constrained in this way, so that the individual matrices $A_j$
can (and will!) still belong to the complexified Lie algebra $sl(2,\C)$.

The conditions needed to make $g$ an element of a coset space 
rather than a group are more subtle and require some explanation.
An essential observation at this point is that, for symmetric $g(\x,\xb)$,
one can prove from \Ref{ls} that the matrix
\be
\M(\t;\x,\xb):= \Psi^t \bigg(\f{1}{\t};\x,\xb \bigg) g^{-1}(\x,\xb)
     \Psi(\t;\x,\xb)   
\la{BMmon1} \ee
is annihilated by the operators \Ref{cd} 
\be \f{d\M}{d\x} = \f{d\M}{d\xb} = 0 \la{BMmon2} \ee
and therefore depends only on the constant spectral parameter, i.e.
\be \M \big( \t(w;\x,\xb); \x,\xb \big) \equiv  \M (w) \la{BMmon3} \ee
(by a slight abuse of notation, we write $\M$ on the r.h.s., too).
{}From the invariance of $w$, and hence of $\M (w)$, under the involution
$\t \ra \t^{-1}$ we immediately obtain
\be
\M^t (w) = \M (w) \la{BMmon4}  \ee
The constancy of the matrix $\M$ (as a function of the coordinates)
was already noticed in \c{BZ}; in \c{BM}, it played an important 
role in reaching a systematic understanding of axisymmetric stationary
solutions of Einstein's equations. $\M(w)$ is called  
``monodromy matrix" in \c{BM}, but obviously it must not 
be confused with the matrices $M_j$ defined in \Ref{monodromy}. 
In both \c{BZ} and \c{BM} the problem of finding solutions to 
Einstein's equations is reformulated as a Riemann Hilbert problem
by reducing \Ref{ls} to the factorization problem \Ref{BMmon1}.
In the description of \c{BM}, $\M =\I$ would correspond to 
the trivial vacuum solution (i.e. Minkowski space), while 
non-trivial solutions are characterized by a non-constant $\M(w)$. 
By contrast, in \c{BZ} $\Psi$ is fixed by demanding
$ \M (w) \equiv \I $ for {\em all} solutions. 
These two possibilities therefore 
correspond to two different descriptions of the 
same solution of Einstein's equation\footnote{The precise
formula can be worked out by factoring the 
``monodromy matrix" of \c{BM} as $ \M_{BM} (w) = S^t(w) S(w)$.
The linear system functions of \c{BM} and \c{BZ} are then related by 
$$ \Psi_{BM} (\t) = \Psi_{BZ} (\t) S\big(w(\t;\x,\xb)\big)$$
This formula also explains why multisoliton solutions correspond
to rational $\Psi(\t)$ in \c{BM}, whereas square root branch
cuts (and hence half integer eigenvalues of $A_j$) appear 
in the description of \c{BZ}.}. 
It is crucial that the asymptotic expansion \Ref{de} is compatible with
the $(\x,\xb)$-independence of the matrix \Ref{BMmon1} only if $\M$
is $w$-independent. For this reason we will adopt the prescription of
\c{BZ} rather than that of \c{BM} in the rest of this paper (this also 
relieves us of the need to worry about a new name for $\M(w)$).

The main idea is now to turn the above statement around and to link 
the desired symmetry of $g$ to the constancy of \Ref{BMmon1}.

\begin{Theorem}
Suppose that in addition to the conditions of Thm.~\r{main} 
function $\Psi$ satisfies
\be
\P(-\bar{\t}) = \overline{\P (\t)}  \la{rc} \ee
and 
\be
\P^t\Big(\f{1}{\t}\Big) g^{-1} \P(\t) =  g_\i
\la{sym} \ee
Then the constants of integration in (\r{h}) and (\r{g}) 
may be chosen such that $g(\x,\xb)$ is symmetric 
and $h (\x,\xb) \in {\bf R}$; in particular, \Ref{sym} implies
$g_\i\in \coset$.) \la{real}
\end{Theorem}

{\it Proof.} From \Ref{rc} we immediately obtain 
\be
\overline{A(\t )} = -  A(-\overline{\t})  \la{Areality}
\ee
so we have in particular $A(1) = - {A(-1)}$; the reality condition
$g(\x,\xb)\in SL(2,\R) $ then follows from \Ref{nk}. Setting $\t = \i$
in \Ref{sym} and making use of the assumed asymptotic properties
\Ref{i} and \Ref{0} of $\Psi$ 
we get $ g =g^t$; on the other hand, taking $\t =0$ we obtain
$g_\i^t =g_\i$. $\Box$

In the form stated above, the conditions suffer from
the drawback that the solution $g(\x,\xb)$ of the Ernst equation 
appears explicitly in \Ref{sym}, so the coset property  
can only be verified a posteriori, i.e. after the solution has
already been constructed. As a first step towards the  
elimination of $g$ we reformulate the relevant conditions 
in terms of the poles $\t_j$ and the matrices  $A_j$.

\begin{Lemma}
Theorem \r{real} remains valid if we require $\Ai =0$ and
replace conditions (\r{rc}) and (\r{sym}) by the invariance of 
the set $\big\{ A_j , \t_j \big\}$ with respect to the involutions
\be
 A_j  \rightarrow \overline {A_j}
\;\;\;,\;\;\;\; \t_j\rightarrow -\bar{\t}_j
\la{i1}\ee 
\be
 A^t_{j} \rightarrow  -  g^{-1} A_j g \;\;\;\;, \;\;\;
\t_{j}\rightarrow\f{1}{\t_j}
\la{mon1}
\ee
\la{l1}
\end{Lemma}
{\it Proof.} Clearly, the first condition is equivalent to \Ref{Areality}
by \Ref{de} (see also \Ref{cur}).
To prove the second part of the lemma, we differentiate
both sides of \Ref{sym} with respect to $\t$; noticing that the r.h.s. 
gives zero a little algebra leads to
\be
\f{1}{\t^2} A^t \bigg( \f{1}{\t} \bigg) = g^{-1} A(\t) g 
\la{step1}  \ee
or
\be
\f{1}{\t} \sum_{j=1}^N \f{A_j^t}{\t^{-1} - \t_j} =
 \t \sum_{j=1}^N \f{g^{-1} A_j g}{\t -\t_j}  \la{step2} \ee
Now performing the substitution \Ref{mon1} on the l.h.s.
we arrive at
\be
\sum_{j=1}^N \t \f{A_j}{\t-\t_j} = \sum_{j=1}^N \t_j \f{A_j}{\t -\t_j}
\la{step3} \ee
which is indeed fulfilled provided that $\Ai = \sum A_j =0$. $\Box$ 

The coset condition \Ref{mon1} can be satisfied by taking $N=2n$ 
and assuming
\be
\t_{j+n}=\f{1}{\t_j} \;\;\;\;\;\;\; A^t_{j+n}=  -  g^{-1} A_j g  
\la{cosetone} \ee
While the involution $\t \rightarrow \t^{-1}$ has only two
fixed points at $\t=\pm 1$, the anti-involution 
$\t\rightarrow -\bar{\t}$ leaves all points
$\t_j$ fixed for which $w_j\in {\bf R}$.
Thus \Ref{i1} implies that $n=m+2l$ and that 
for $j=1,...,m$ and $j=n+1,...,n+m$ 
\be
\t_j=-\bar{\t_j}\;\;\;\;\;\; A_j = \overline{A_j} \;\;\;\;\;
\la{tjm}\ee
whereas, for $j=m+1,...,m+l$ and $j=n+m+1,...,n+m+l$,
\be
\t_j=-\overline{\t_{j+l}}\;\;\;\;\;\; 
A_j = \overline{A_{j+l}}
\la{tjl}\ee
Thus for \Ref{tjm} we have $w_j \in \R$ and $A_j \in sl(2,\R)$
whereas for \Ref{tjl} $w_j$ is complex and we have $A_j \in sl(2,\C)$. 

The complete elimination of $g(\x,\xb)$ is achieved by reformulating 
the above constraints in terms of the monodromy matrices and the 
$\tau$-function. It is straightforward to deduce from \Ref{RH}
and \Ref{sym} that
%\be
%\f{1}{\t^2} \Big( \Psi^{-1} \Psi_\t \Big)^t \bigg( \f{1}{\t} \bigg)
% = g_\i \big( \Psi^{-1} \Psi_\t \big) (\t) g_\i^{-1} 
%\la{step4}   \ee
%With the help of \Ref{monodromy} we thus obtain the following
%relations for the monodromy matrices
\be  M_{j+n}^t = g_\i M_j^{-1} g_\i^{-1} \la{step5} \ee
which do not contain $g(\x,\xb)$ anymore. By the one-to-one
correspondence between the solutions of \Ref{ee} and the 
monodromy data \c{Jimbo}, we therefore have
\begin{Theorem}
Let $\Psi$ obey the reality condition \Ref{rc} and let
the monodromy matrices satisfy the relations \Ref{step5}.
Then the constants of integration in \Ref{nk} and \Ref{h} may be 
chosen in such a way that $g(\x,\xb)$ solves the Ernst equation \Ref{ee}, 
is real and symmetric, and the conformal factor $h$ is real.
\la{lem1}
\end{Theorem}
Finally, we give the necessary condition for the fulfilment
of the constraints in terms of the $\tau$-function.
\begin{Corollary}
Suppose that the same conditions as in Lemma~\r{l1} hold. Then
\be
\tau\left(\f{1}{\t_1},...,\f{1}{\t_N}\right)= \tau(\t_1,...,\t_N)
\la{tau0}\ee
\la{lem}
\end{Corollary}
{\it Proof.} 
From \Ref{om0} it is obvious that the $\tau$-function depends 
only on the traces of products of the matrices $A_j$. 
Then \Ref{sym} is an immediate
consequence of the invariance under \Ref{mon1}. $\Box$.

The symmetry and reality conditions as well as the regularity 
properties of $\Psi$ are preserved by the Ehlers group $SL(2,\R)$ 
\c{Exact}). An Ehlers transformation is
characterized by a matrix $Q\in SL(2,\R)$ and acts on $\Psi$ as 
\be
\P\ra\Psit=Q^t \Psi Q
\la{Ehlers1}\ee
{}From \Ref{A(t)} we infer that \Ref{Ehlers1} induces the
following transformation on the residue matrices
\be
A_j \ra \tilde{A}_j= Q^t A_j Q^{t\; -1} 
\la{Ehlers2} \ee
It is known that the Ehlers group admits an infinite extension,
the so-called Geroch group \c{Ger}, which also acts on the 
space of axisymmetric stationary solutions. Save for some
scattered remarks we will not consider this group here.
It is efficiently described in the 
supergravity inspired approach of \c{Jul,BM,Nic} which appears 
to be best suited for understanding the underlying 
``hidden symmetries" in a systematic fashion.

\subsection{``Two-time" Hamiltonian formalism} 
We will now present the canonical formulation of the results
described in the foregoing sections. For this purpose we adopt
a ``two-time" Hamiltonian formalism with the two ``times" corresponding 
to the lightcone coordinates $\x$ and $\xb$. One major advantage of this
procedure is that the quantum theory is manifestly covariant
under $2d$ coordinate transformations, a feature which is far from
obvious (and possibly not even true) for the ADM formulation 
of canonical quantum gravity (see e.g. \c{multi-time}
for a recent discussion). In the case
at hand the two-time Hamiltonian structure giving rise to the  
equations of motion (\r{1}) is summarized in 

\begin{Theorem}
The system (\r{1}) is a ``two-time"
Hamiltonian system with respect to the Lie-Poisson bracket
\be
 \left\{A(\t)  \stackrel{\otimes}{,}  A(\mu)\right\} =
\Big[ r(\t-\mu)\, ,\, A(\t)\otimes \I + \I \otimes A(\mu)\Big]
\la{pb}\ee
where $A(\t)\equiv \P_{\t}\P^{-1}$ and the classical rational
$R$-matrix $r(\t)$ is equal to $\Pi/\t$ with $\Pi$ the
permutation operator in $\C^2\times \C^2$:
$$\Pi=\left(\baa{cccc}1\;\;0\;\;0\;\;0\\
                     0\;\;0\;\;1\;\;0\\
                     0\;\;1\;\;0\;\;0\\
                     0\;\;0\;\;0\;\;1\eaa\right) $$
The dynamics of the physical fields in the $\x$
and $\xb$-directions are governed by the ``matter Hamiltonians"
\begin{eqnarray}
H^{(\x)} &:= &\f{1}{\x-\xb} {\rm tr} \, A^2 (1)
 = \f{1}{\x - \xb} \sum_{j,k=1}^N
\f{\tr (A_j A_k)}{(1-\t_j)(1-\t_k)}  \nonumber \\
H^{(\xb)} & :=& \f{1}{\xb-\x}{\rm tr} \, A^2 (-1)
 = \f{1}{\xb - \x} \sum_{j,k=1}^N
\f{\tr (A_j A_k)}{(1+\t_j)(1+\t_k)}   \la{Ham}
\end{eqnarray}
and the respective
flows generated by $\Hx$ and $\Hxb$ commute, i.e. $\{\Hx,\Hxb\}=0$.
\la{Poisson}
\end{Theorem}

{\it Proof.} The main statement can be verified by direct calculation. 
Commutativity of the Hamiltonians follows from the more general relation
\be \Big\{ \tr A^2(\t)  \, , \, \tr A^2(\mu) \Big\} = 0
\la{commute} \ee
which is valid for arbitrary $\t$ and $\mu$. 
The commutativity of the flows generated by
by $\Hx$ and $\Hxb$ is equivalent to the decoupling of
the classical equations of motion in \Ref{1}, and may be viewed
as a direct consequence of the compatibility of the 
system (\r{1}) (cf. Thm. \r{commut}).
Observe that we have $(\Hx )^\dagger = \Hxb$. 
$\Box$

For the benefit of readers not familiar with the above tensor product
notation (see \c{FT, Korep} for details), we spell out these brackets
once more with matrix indices $\a,\b,\dots =1,2$ indicated explicitly.
Setting
$$ \Big( A(\t) \otimes A(\mu)\Big)_{\a \b,\g \d} :=
  A(\t)_{\a \g} A(\mu)_{\b\d}
$$
and 
$$\Pi_{\a \b,\g \d} = \delta_{\a\d} \delta_{\g \b}$$
the Poisson brackets \Ref{pb} are equivalent to
$$
\big\{ A_{\a\b} (\t ) , A_{\g\d} (\mu) \big\} =
\frac{1}{\t - \mu} \bigg( \delta_{\a\d} 
     \big( A(\t)- A(\mu )\big)_{\g\b} - \delta_{\g\b}
     \big( A(\t)- A(\mu )\big)_{\a\d}  \bigg) .
$$  
Defining
\be
A_{\a \b} \equiv A_j^a t^a_{\a\b} \la{Aab} \ee
where $t^a$ are the generators (so far of $SL(2,\C)$)
and inserting \Ref{A(t)} into \Ref{pb}, we get
\be  
 \big\{ A_j \stackrel{\otimes}{,} A_k \big\} = 
 \delta_{jk} \Big[ \Pi \, , \, A_j \otimes \I \Big]
\la{pbAjAk}  \ee
or, equivalently,
$$\{A_j^a,\;A_k^b\}=\delta_{jk}{f^{ab}}_c A_j^c $$
where ${f^{ab}}_c$ are the structure constants of $SL(2,\C)$.

With (\r{Ham}) we can thus reexpress the equations for the conformal
factor (\r{h1}) in the form
\be
\cC^{(\x)} := - \k_{\x} + \Hx \approx 0 \;\;\;\;\; , \;\;\;\;\;
\cC^{(\xb)} := - \k_{\xb} + \Hxb \approx 0      \la{WDW}
\ee
Thus, in accordance with the general theory of constraints \c{Dirac},
we shall from now on regard (\r{WDW}) as constraints \`a la Dirac
rather than merely as equations determining the conformal factor
(accordingly, $\approx$ means ``weakly zero"). 
This interpretation is appropriate for generally covariant theories
where the local (gauge) invariances give rise to canonical constraints
whose ``matter parts" are just the conventional Hamiltonians \Ref{Ham}. 
More precisely, the constraints \Ref{WDW}
express the invariance of the theory under local translations
in $\x$ and $\xb$, respectively; as such they are linear
combinations of the WDW and diffeomorphism constraints 
corresponding to the invariance of the theory with respect to
local translations in ``time" $\rho$ and ``space" $x$.
As before we have a reality condition
$(\Cx )^\dagger = \Cxb$ for the constraints.

Interpreting \Ref{WDW} as canonical constraints requires that 
we enlarge the phase space so as to account for the gravitational 
degrees of freedom (the conformal factor and the dilaton). 
Their canonical brackets are given by
\be
\{ \x, 2k_\x \} = \{ \xb, 2k_\xb  \} =1\;\;\;\;\;\;\;\;\;\;
\{ \xb,  2k_\x  \} = \{ \x, 2k_\xb \} =0
\la{pbh}\ee
(Strictly speaking, the derivation of these brackets would
require that we undo the choice of Weyl coordinates, 
on which (\r{metric}) and (\r{ee}) are based, but we will 
skip this step here.) Use of \Ref{pbh} and some
further computation then show that the constraints commute
like the matter hamiltonians in terms of which they are defined
\be
\{ \cC^{(\x)} , \cC^{(\xb)} \} = 0  \la{constraint1}
\ee

A noteworthy feature of the new formulation
is that the dimension of the system has been effectively reduced from
two to one by trading the ``space" variable $x$ and the
``time" variable $\rho$ for two ``time" variables $\x$ and $\xb$.
We can thus regard the spectral parameter
currents $A(\gamma )$ at a fixed but arbitrarily chosen
base point $(x_0, \rho_0 )$ as the fundamental canonical variables.
In other words, instead of considering phase space variables 
depending on the space coordinates, we now take them to depend 
on the spectral parameter. Since this point can be chosen at will,
this formulation manifestly preserves $2d$ covariance.
The ``time evolutions" of any phase space function
$F\big( \{A_j\};\x,\xb,k_\xb,k_\x \big)$ are then
generated as usual by commutation with the constraints
$\Cx$ and $\Cxb$, i.e.
\be
\f{d F}{d\x} = \{ \Cx , F \}  \;\;\; , \;\;\; 
\f{d F}{d\xb} = \{\Cxb , F \}  \la{evolve}
\ee
It is important that the derivatives appearing here are 
{\em total} derivatives with respect to $\x$ and $\xb$,
with the first term of $\Cx$ or $\Cxb$ generating the partial 
derivatives with respect to the coordinates, and the second term
taking care of the $(\x,\xb)$-dependence of $A_j$.
Altogether the action of the constraints
on any phase space function is thus simply given by the operators
\be
\Cx \cong \f{d}{d\x} \;\;\; , \;\;\; \Cxb \cong \f{d}{d\xb}
\la{C=d/dx} \ee
As we will see this remains true for the quantized theory 
where the constraints become operators acting on a Hilbert space
of wave functionals.

Observables in the sense of Dirac are by definition all 
those functionals $\cO$ on phase space which weakly 
commute with the constraints $\Cx$ and $\Cxb$
but do not vanish on the constraint hypersurface 
$\Cx = \Cxb = 0$, i.e.
\be
\{ \Cx , \cO \} \approx 0 \;\;\; , \;\;\; \{ \Cxb , \cO \} \approx 0 
\la{observable} \ee
By \Ref{C=d/dx} the observables are independent of the 
coordinates and therefore highly non-local objects as one
would expect on general grounds \c{QG,Ashtekar}. It is easy to see 
that our model admits an infinite number of independent observables.
First of all, the parameters $w_1,...,w_N$ trivially belong
to this class since they commute with everything. Secondly, and 
more importantly, the monodromies $M_1,...,M_N$ defined in \Ref{mon} are 
also observables for arbitrary $N$. This fact is obvious from
\Ref{monconstant}, but can also be rephrased in canonical language.
Namely, as functionals on phase space, the monodromy matrices 
depend both on $\{A_j\}$ and $\{\t_j\}$; furthermore, $M_j$ does {\em not}
commute with the matter Hamiltonians $\Hx$ and $\Hxb$
but only with the
full constraints $\Cx$ and $\Cxb$ of \Ref{WDW}; for instance,
$$ \f{d M_j}{d\x} =
\{\Cx \, , \, M_j\}=-\{2 k_{\x},\x \}\f{\p M_j}{\p\x }+ \{\Hx, M_j\}=
\f{\p M_j}{\p\x} +\{\Hx , M_j\} =0 $$
We here recognize an important difference between dimensionally
reduced gravity and the corresponding flat space $\s$-models, 
where $\t$ would be coordinate independent and the traces 
$\tr A(\t)^2$ would already be constants of motion by \Ref{commute}.

All observables can be generated from the set 
\be
\obs := \big\{w_1,...,w_N; M_1,...,M_N \big\}
\la{sbo}\ee
by taking arbitrary products and linear combinations. In this sense
$\obs$ constitutes a complete set of classical (and quantum) observables
for arbitrary $N$. These are the conserved ``non-local charges"
of dimensionally reduced gravity. The monodromies are not 
easy to handle at the canonical level because of their non-local 
dependence on  $A(\t)$ \c{Buckow}. For this reason 
we shall mostly deal with the restricted set of observables 
\be
\sbo :=
\Big\{ w_1,...,w_N\, ; \, \tr A_1^2,\dots,\;\tr A_N^2\, ; \, A_{\i} \Big\}
\subset \obs
\la{obs2}
\ee
We include $\Ai$ (cf. Lemma~\ref{intmot}) here to keep the discussion
as general as possible. If $\Psi$ is regular
at infinity the condition $\Ai = 0$ should be
treated as a first class constraint since this matrix
obviously closes into an $SL(2,\R)$ algebra and its entries 
are integrals of motion. The meaning of $\Ai$ is further clarified by
the following

\begin{Theorem}
The matrix elements of $A_{\i}$ are the canonical
generators of the group of Ehlers transformations with respect 
to the Poisson structure \Ref{pb}.
\end{Theorem}
{\it Proof.} By \Ref{Ehlers1}, an infinitesimal Ehlers transformation
with parameter $\epsilon_{\a\b}$ acts on $A_j$ as
$$ \d A_j = [ A_j , \epsilon ]   $$
On the other hand, we have
$$ \big\{ \epsilon_{\g \d} (A_\i)_{\d \g} \, , \, (A_j)_{\a\b} \big\}
   = [ A_j , \epsilon ]_{\a\b}  $$
by \Ref{pbAjAk}.
$\Box$  

At this point the following comments concerning the status 
of the constraints \Ref{Areality} and \Ref{tau0} are in order.
In the sequel we shall mainly deal with the case $\t_j=-\bar{\t}_j$
(i.e. set $m=n$ in \Ref{tjm}). Then condition \Ref{Areality} 
just means that $A_j\in sl(2,\R)$ and the Poisson bracket \Ref{pbAjAk} 
is the standard Kirillov-Konstant bracket for $sl(2,\R)$.
The asymptotic regularity condition $\Ai =0$ should be
considered as a part of the coset constraints. It ensures the
symmetry of $\Psi$ with respect to the involution
$\t\rightarrow \t^{-1}$; regularity of $\Psi$ at $\t=\i$ then
follows from the assumed regularity of $\Psi$ at $\t =0$.
However, the constraint \Ref{tau0} is not an ordinary phase space
constraint since it relates the  phase space
variables at the different ``times" $\t_j$. 

Let us mention two alternative ways of writing the brackets (\r{pb}):
\begin{itemize}
\item
The total phase space of the theory may also be parametrized by the
coefficients of the expansion of $A(\t)$ at $\t=\i$:
\be
A(\t)=\sum_{k=1}^{\i} \t^{-k}\tilde{A}_k 
\;\;\;\;\;\;\;\;\;\tilde{A}_k\equiv\sum_{j=1}^{N} A_j\t_j^{k-1}
\la{infty}\ee
Assuming the total phase space to be
$${\cal H}=\big\{\tilde{A}_k\;|\;k=1,2,...\big\}$$
and substituting (\r{infty}) into (\r{pb}) we get
\be
\{\tilde{A}_j^a,\tilde{A}_k^b\}= {f^{ab}}_c \tilde{A}_{j+k}^c\;\;\;\;\;j,k>1 ,
\la{pb0}\ee
i.e. ``half" of the current algebra $A_1^{(1)} \equiv \widehat{sl(2,\C)}$. 

For the isomonodromic sectors with finite $N$, the currents
$\tilde{A}_j$ are clearly dependent. For $N\rightarrow\i$, on the other 
hand, they can be considered as independent canonical variables; for 
arbitrary non-isomonodromic solutions the natural way to generalize
our present construction might thus be in terms of the variables 
$\tilde{A}_j$ together with (\r{pb0}). 
The Ehlers charge coincides with the first term of the expansion
\Ref{infty}, i.e. $-A_{\i} \equiv \tilde{A}_1$.

\item
For the infinite set of currents 
\be
J_{m,n} \equiv \sum_{j=1}^N \f{A_j}{(1-\t_j)^m (1+\t_j)^n }
\;\;\;\;\;;\;\;\;
J_{1,0}=\f{\x-\xb}{2} g_{\x} g^{-1}
\;\;\;\;\;\;\;\;
J_{0,1}=\f{\xb-\x}{2} g_{\xb} g^{-1}
\la{hc}\ee
we have the ``current algebra"
\be
\{J_{m,n}^a , J_{m',n'}^b\} = {f^{ab}}_c J_{m+m',n+n'}^c
\la{pb1}\ee
Now  $-A_{\i}$ coincides with $J_{0,0}$.

\end{itemize} 

The Poisson structure \Ref{pb} also appears in 
Chern-Simons theory where one starts from the connection
$$ {\cal A}:= A_1(\t,\bar{\t})d\t + A_2(\t,\bar{\t})d\bar{\t} $$
with the bracket
$$\{A_1^a (\t,\bar{\t}), A_2^b(\mu,\bar{\mu})\} = \delta^{ab}
\delta^2 (\t-\mu) $$
Imposing the flatness condition
$$A_{1\bar{\t}}-A_{2\t}+[A_1, A_2]=0 $$
and choosing a holomorphic gauge $A_2=0$ (see \c{FR}), 
one can derive the bracket \Ref{pb} as the Dirac bracket for 
the remaining component $A(\t)\equiv A_1(\t)$,  at least for the
punctured sphere. This link is explained
in more detail in \c{KS,Buckow}. 

To conclude this section we would like to briefly comment on the relation
between the new Hamiltonian formulation and the
conventional one based on the use of one space and one time variable, 
where $g(\x,\xb)$ would be treated as a quantum field
coupled to $2d$ gravity and a dilaton.
It will be sufficient to explain the differences
for the principal chiral model (see e.g. \c{deVega} for a description of
the corresponding flat space model). There the main objects of interest
are the $x$-dependent currents at fixed ``time" $\rho$
\be
   J_\rho = \rho g_{\rho} g^{-1} \;\;\;\;\;{\rm and}\;\;\;\;\;
   J_x = \rho g_x g^{-1}
\la{current}
\ee
(which are obviously related to the currents $J_{0,1}$ and
$J_{1,0}$ from \Ref{hc})
and their equal time Poisson brackets, which read
\begin{eqnarray}
\{ J_\rho (x) \stackrel{\otimes}{,} J_\rho (y) \} & = & \rho
\Big[ \Pi \, , \, J_\rho (x)\otimes \I \Big] \, \delta (x-y) \nonumber \\
\{ J_\rho (x) \stackrel{\otimes}{,} J_x (y) \} &=& \rho
\Big[ \Pi \, , \, J_x (x)\otimes \I  \Big] \, \delta (x-y)
+ \rho \Pi \, \delta'(x-y) \nonumber \\
\{ J_x (x) \stackrel{\otimes}{,} J_x (y) \} &=& 0
\la{PB3}
\end{eqnarray}
(for coset space $\s$-models the matrix $\Pi$ would not be 
independent of the fields).
The Hamiltonian determining the evolution in the $\rho$-direction is
\be
H=\int\rho^{-1}{\rm tr} (J_\rho^2 - J_x^2) dx
\la{H}
\ee
The non-ultralocal $\delta'$-term in (\r{PB3}) has been
a notorious source of trouble, leading to
irresoluble ambiguities in the Poisson brackets of certain integrated
phase space quantities \cite{deVega,Faddeev,DNN}.
It also represents a serious obstacle towards
the application of standard quantization techniques \c{FT, Korep}. 
By contrast, our ``two-time" formalism sidesteps this difficulty, 
as the troublesome non-ultralocal term has disappeared. 
Beside yielding the same canonical equations 
of motion as the usual approach, the (real)
Hamiltonian defining the evolution in $\rho$-direction
in our formalism is equal to (recall from \Ref{gaugefix}
that $\rho = {\rm Im} \, \xi$) 
\be
H^{(\rho)}= \f{i}{2}(\Hx-\Hxb) =-\f{\rho}{4}
   \Big({\rm tr}(g_{\x}g^{-1})^2
   +{\rm tr} (g_{\xb}g^{-1})^2 \Big) =
\f{\rho}{2}{\rm tr} \Big( (g_\rho g^{-1})^2 - (g_x g^{-1})^2 \Big)
\la{Hrho}
\ee
and therefore agrees with the Hamiltonian {\em density} (\r{H})
of the usual approach. Let us note that the
minus sign in front of $J_x^2$ is simply due to the ``Wick rotation" 
of the $1+1$ metric to a $2+0$ metric; the apparent lower unboundedness
of the Hamiltonian \Ref{Hrho} is therefore spurious.

\subsection{Relation to the Schlesinger equations}

Our equations \Ref{1} are closely related to the so-called 
Schlesinger equations \c{Sch} which play an important role in
the theory of integrable systems \c{Jimbo}. To exhibit the relation,
let us consider $\t_j,\;j=1,...,N$ as independent deformation
parameters and suppose that the monodromy data $\{T_j\;,C_j\}$
are $\t_j$-independent.
%this means that in comparison with our previous
%treatment we have to add a formal condition of $\l_j$-independence
%of the monodromy data. 
Instead of \Ref{de} and \Ref{ls}, we 
would then get the following deformation equations in $\t_j$
\be
\f{\p \P}{\p \t_j}=-\f{A_j}{\t-\t_j}\P\;\;\;\;,\;\;\;\;j=1,...,N
\la{ls1}\ee
Demanding compatibility of (\r{ls1}) and (\r{de}) we arrive at the
classical Schlesinger equations \c{Jimbo}
\be
\f{\p A_{j}}{\p\t_k}
            =\f{[A_j,A_k]}{\t_j-\t_k}\;\;\;\;(k\neq j)\;\;\;\; ; \;\;
\;\;\;\;
\f{\p A_{j}}{\p\t_j}
            =- \sum_{i\neq j}
\f{[A_j,A_i]}{\t_j-\t_i}
\la{Sch}\ee
The system (\r{Sch}) is an $N$-time Hamiltonian system with respect to
the Poisson structure (\r{pb}) \c{Jimbo} with  times $\t_j$ for the
Hamiltonians 
\be
H_j=\sum_{i\neq j} \f{\tr (A_i A_j)}{\t_j-\t_i}\;\;,\;\;\;\;\;
j=1,...,N
\la{Hj}\ee
The Hamiltonians $H_j$ mutually commute and
can alternatively be obtained from
\be
H_j ={\rm res} \Big|_{\t=\t_j} \tr A^2(\t)
\la{Sug1}\ee
The $\tau$-function (\r{tau}) is the generating function for Hamiltonians
$H_j$ in the sense that
\be
\frac{\partial \tau}{\partial \t_j}= H_j \tau
\la{gen}\ee

Now we are in position to formulate the theorem relating the Schlesinger
equations (\r{Sch}) to our deformation equations (\r{1}).

\begin{Theorem}
Let the functions $A_j(\{\t_k\})$, $j=1,...,N$  solve the Schlesinger
equations (\r{Sch}) and obey the constraint 
(\r{regularity}). Furthermore,
let the variables $\t_j$ depend on $(\x,\xb)$ according to
(\r{gamma}), i.e. $\t_j=\t (\l_j;\x,\xb)$. Then the residue functions
$A_j(\{\t_k(\x,\xb)\})$ satisfy equations (\r{1}).
\la{link}
\end{Theorem}
The proof is straightforward. Using (\r{Sch}), (\r{pe}), we get,
for example,
$$ A_{j\x}= \sum_{k}\f{\p A_j}{\p\t_k} \t_{k\x} =
\f{1}{\x-\xb}\sum_{k\neq j}\f{[A_j,A_k]}{\t_j-\t_k}\left\{
\t_k\f{1+\t_k}{1-\t_k} -\t_j\f{1+\t_j}{1-\t_j}\right\}  $$
\be
=\f{2}{\x-\xb}\sum_{k\neq j}\f{[A_k,A_j]}{(1-\t_j)(1-\t_k)} +
\f{1}{\x-\xb}\Big[ A_j,\sum_k A_k  \Big]
\la{cal}\ee
The constraint  (\r{regularity}) 
eliminates the last term in (\r{cal}).$\Box$

To clarify the link between Hamiltonians $\Hx$, $\Hxb$ (\r{Ham}) and the
Hamiltonians $H_j$ (\r{Hj}), we note that the evolution in $\x$-direction
of an arbitrary solution of (\r{Sch}) is given by the Hamiltonian
$$\sum_j H_j\t_{j\x} = \f{1}{\x-\xb}
\sum_{k\neq j} \f{\tr (A_j A_k)}{(1-\t_j)(1-\t_k)} -
\f{1}{2(\x-\xb)}\sum_{k\neq j} \tr (A_j A_k)  $$
Using \Ref{trAjAk} and comparing this result with \Ref{Ham} we get
\be
\Hx =\sum_j H_j\t_{j\x}
-\f{1}{\x-\xb}\sum_j \tr A_j^2 \left\{\f{1}{2}-\f{1}{(1-\t_j)^2}
\right\} +\f{1}{2(\x-\xb)}\tr (\Ai)^2 
\la{calc1}\ee
Since the terms containing $\tr A_j^2$ commute with all $A_k$ by virtue of
(\r{pb}), they do not give any contribution to the equations
of motion, and thus can be interpreted as contributing to the ``vacuum
energy" only (the last term on the r.h.s. obviously vanishes for 
asymptotically flat solutions). 
Upon quantization the ``vacuum energy terms"
turn into Casimir operators, and do contribute to the 
wave function via an explicitly computable phase factor. 

The Schlesinger equations together with \Ref{gamma} imply that the
dependence of $A_j$ (and in fact any phase space variable depending
on the $A_j$'s) on the parameters $w_j$ is governed 
by the mutually commuting Hamiltonians
\be
\Hwj=\f{\partial \t_j}{\partial w_j} H_j
\la{Hwj}\ee
since
\be 
\f{\partial A_j}{\partial w_k} = \Big\{ H^{(w_k)} \, , \, A_j \Big\}
\la{dAdwk}  \ee 
Therefore the Hamiltonian $\Hwj$ \Ref{Hwj} can be interpreted as
a generator of translations in the variable $w_j$, as it 
moves the position of the $j$-th singularity.
The Hamiltonians \Ref{Hwj} also commute with the total Hamiltonians 
$\Hx$ and $\Hxb$ provided \Ref{regularity} is satisfied. 
In fact, \Ref{dAdwk} means, that, in analogy to evolution in $\x$
and $\xb$ directions, the proper treatment of the evolution in
$w_j$ direction is the following: the canonically conjugate variable to
$w_j$ is $\k_{w_j}$; then equation $\k_{w_j}- H^{(w_j)}=0$
should be treated as a constraint in analogy with the constraints 
$\Cx$ and $\Cxb$ \Ref{WDW}.

\section{Quantization}
\setcounter{equation}{0}

\subsection{Commutation relations and  Bethe ansatz}

To quantize the model, we replace the Poisson
brackets (\r{pb}) by commutators in the usual fashion:
\be
[A(\t)\stackrel{\otimes}{,}A(\mu)]
= i\hbar \Big[r(\t-\mu)\;,\;A(\t)\otimes \I + \I\otimes A(\mu)\Big]
\la{CR}\ee
The entries of the matrix of $A(\t )$ thus become operators acting on a
Hilbert space to be specified below;
note that on the l.h.s. of (\r{CR}),
we have a commutator of operators in Hilbert space
whereas on the r.h.s. we have a commutator of ordinary matrices.
This means in particular that the expansion (\r{de}) is no longer valid
as an operator statement, but must be reinterpreted as a property of
the states on which $A(\t)$ acts. We write
\be
A(\t)\equiv \f{i \hbar}{2}\pmatrix{\h(\t) & 2\e(\t) \cr 2\fg(\t) & -\h(\t) \cr}
\la{Aquant}\ee
The reality constraint \Ref{Areality} translates into
\be
\h(\t)^{\dagger} = \h (-\overline{\t}) \;\;\; , \;\;\;
\e(\t)^{\dagger} = \e(-\overline{\t})  \;\;\; , \;\;\;
\fg(\t)^{\dagger} = \fg(-\overline{\t})  
\la{hermiticity}
\ee
\Ref{CR} and \Ref{Aquant} yield the following commutation relations
for the operators $\h(\t), \e (\t)$ and $\fg(\t)$
\begin{eqnarray}
[\h(\t), \e(\t')] &=& - \frac{2}{\t-\t'} \Big( \e (\t)- \e (\t') \Big) 
 \nonumber \\ {}
[\h(\t), \fg(\t')] &=& \frac{2}{\t-\t'} \Big( \fg(\t)-\fg(\t') \Big)
 \nonumber \\ {}
[\e(\t) , \fg(\t')] &=& - \frac{1}{(\t-\t')}\Big(\h(\t)-\h(\t') \Big)
\la{Chevalley1}
\end{eqnarray}
with all other commutators equal to zero. For 
coincident arguments we have, for instance,
$$
[\h(\t) , \e(\t) ] = -2 \frac{d \e (\t)}{d\t}
$$

The Hamiltonians (\r{Ham}) remain unchanged; they can be written out
more explicitly in terms of the matrix elements of $A(\t)$ and
thereby cast into a form reminiscent of the Sugawara construction. 
Explicitly,
\be
\Hx=  \f{1}{\x-\xb}\tr\Big( A^2(1)\Big)\equiv
-\f{\hbar^2}{\x-\xb} \Big( \ft12 \h(1)\h(1) + \e(1)\fg(1) 
+ \fg(1)\e(1)\Big)
\la{Sug}\ee
\be
\Hxb =  - \f{1}{\x-\xb}\tr\Big( A^2(-1)\Big)\equiv
 \f{\hbar^2}{\x-\xb}  \Big( \ft12 \h(-1)\h(-1) + \e(-1)\fg(-1) 
+ \fg(-1)\e(-1)\Big)
\la{Sugaw}\ee

As already mentioned
we shall restrict ourselves to states on which the operator 
$A(\t)$ may be represented as in \Ref{de} with $\t_j=-\bar{\t}_j$. 
Accordingly, we put
\be
A_j\equiv  \f{i\hbar}{2}
   \pmatrix{ \h_j & 2\e_j \cr 2\fg_j & -\h_j \cr} 
\ee
so that
\be 
\h(\t) = \sum_{j=1}^N \f{\h_j}{\t - \t_j}  \;\;\; , \;\;\;
\e(\t) = \sum_{j=1}^N \f{\e_j}{\t - \t_j}  \;\;\; , \;\;\;
\fg(\t) = \sum_{j=1}^N \f{\fg_j}{\t - \t_j}  \la{eft} 
\ee
The operators $\h_j ,\e_j $ and $\fg_j$ are the anti-hermitian 
Chevalley generators of $SL(2,{\bf R})$ (see App. A)
obeying the standard commutation relations
\be
[\h_j \, , \, \e_j]= 2\e_j\;\;\; , \;\;\; [\h_j\, ,\,\fg_j]= -2\fg_j 
\;\;\; , \;\;\; [\e_j\, , \, \fg_j]= \h_j
\la{CR2}\ee
as a consequence of (\r{Chevalley1}). 

For the explicit construction of solutions it is convenient
to switch from the $SL(2,\R)$ basis to an $SU(1,1)$ basis
in terms of which raising and lowering operators can be defined;
this also facilitates the comparison between the non-compact case 
$G=SU(1,1)$ and the compact case $G=SU(2)$.
The $SU(1,1)$ Chevalley generators are defined by
\be
e_j:=\ft12(-i \h_j + \e_j+\fg_j)\;\;\;\;\;\;
f_j:=\ft12(i\h_j +\e_j+\fg_j)\;\;\;\;\;
h_j:=i(\fg_j-\e_j) \la{SU(1,1)}
\ee
They also obey
\be
[h_j\, , \, e_j]= 2e_j\;\;\; , \;\;\; [h_j\, ,\,f_j]= -2f_j \;\;\; , \;\;\;
[e_j\, ,\,f_j]=  h_j
\la{CR3}\ee
but are no longer anti-hermitean, satisfying instead  
the following hermiticity properties 
\be
h_j^{\dagger} = h_j\;\;\;\;\;\;
e_j^{\dagger}= - f_j
\la{rc0}\ee
For $SU(2)$, the operators are the same as in \Ref{SU(1,1)}
but the hermiticity condition reads
\be
h_j^{\dagger} = h_j\;\;\;\;\;\;
e_j^{\dagger}= + f_j  \la{rcc0}\ee
In both cases we can interpret $e_j$ and $f_j$ as creation and
annihilation operators, respectively (or vice versa) and
diagonalize the operators $h_j$. 

Physical states based on unitary representations of $SU(2)$ or
on the discrete series representations of $SU(1,1)$ always
admit a ``ground state" $\0$ labeled by some analytic
function $p = p(\t )$ subject to the reality condition
$\overline{{p(\t)}} = p(-\bar \t )$. This state is a lowest weight state
in the sense that it is assumed to obey the conditions
\be 
h(\t) \0 = p(\t) \0 \;\;\;\;\;\;{\rm and}\;\;\;\;\;\; 
f(\t) \0 =0  \la{Bethe1} \ee
The classical expansion (\r{de}) corresponds to the special choice
\be
p (\t) =  \sum_{j=1}^N \f{ s_j}{\t - \t_j}
\la{at}
\ee
which is equivalent to $h_j \0 = s_j \0$. The ``excited states"
are obtained by applying the raising operators to $\0$.
More specifically, we define the (off-shell) Bethe states 
for both $SU(2)$ and $SU(1,1)$ by
\be
|p;v_1,...,v_M \rangle := e(v_1)...e(v_M) \0  \la{Bethe}
\ee
where $e(\t)$ is the analog of \Ref{eft}
and the complex parameters $v_i$ are arbitrary at this point.
Thus the operators $e(\t)$ and $f(\t)$ indeed play the role of 
creation and annihilation operators, respectively, as asserted above; 
interchanging them turns the representation ``upside down". 
For $SU(2)$, the possible values for $s_j$ are negative integer 
(since we are working with lowest weight states) and 
the spectrum of $h_j$ (and hence the number of excitations in
\Ref{Bethe}) is bounded with eigenvalues $-|s_j|,-|s_j|+1, \dots, |s_j|$. 
For the discrete representations of $SU(1,1)$, $s_j$ is either 
positive integer with $s_j \geq 2$ and a semi-infinite 
``topless" spectrum $s_j, s_j+1, \dots$, or negative integer 
with $s_j \leq 0$ and a semi-infinite ``bottomless" spectrum with
$s_j, s_j -1, \dots$. The eigenvalue 
of the Casimir operator is always $s_j(s_j-2)$ in these conventions.
For these representations, the physical parameters (masses, etc.) 
are quantized, and we can claim that the quantum
theory is ``more regular" than the classical theory because
the quantization of $s_j$ severely constrains the
types of singularity that can appear in the quantum wave functional.
For the principal and supplementary series of $SU(1,1)$, 
on the other hand, which have no $SU(2)$ analog, 
$s_j$ is a continuous parameter, and there is consequently
no quantization of physical parameters. 
\Ref{at} and the first relation in \Ref{Bethe1} still hold but
the Bethe ansatz \Ref{Bethe} no longer works because 
the corresponding ground state does not exist.

For the full theory we must also quantize the gravitational 
degrees of freedom $\k_{\x}\;,\;\k_{\xb}$ and $\x$, $\xb$, 
keeping in mind that these are really fields in a special
gauge and not just coordinates on the worldsheet. We must therefore replace
the Poisson brackets \Ref{pbh} by commutators and construct an   
operator representation for the gravitational phase space variables. 
{}From the canonical brackets (\r{pbh}) we deduce
$$
[ \x, \k_{\x}  ] = [\xb, \k_{\xb} ] = i\hbar  \;\;\; , \;\;\;
[ \xb, \k_{\x}  ] = [\x, \k_{\xb} ] = [\k_{\x},\k_{\xb}] = 0
$$
To realize these commutation relations we take
\be
\k_{\x} = - i\hbar \f{\p}{\p \x} +f(\x,\xb)\;\;\; , \;\;\;
\k_{\xb} =- i\hbar \f{\p}{\p \xb}+ \bar{f}(\x,\xb)\la{cofa}
\ee
with an arbitrary function $f$ satisfying
$$\f{\p f}{\p\xb}=\f{\p\bar{f}}{\p\x} $$
A different choice of $f$ will result in a renormalization of
the physical states by a function of the coordinates and will not
affect the physical content of the theory; without loss
of generality we can thus set
$$f(\x,\xb)=0$$

The main advantage of \Ref{cofa} is that by
representing $(\x, \xb )$ as multiplication operators
we salvage their interpretation as coordinates; otherwise the spectral
parameter $\t$ would not remain a function but become a non-local
differential operator and thus very awkward to deal with.
It is then obvious that the two equations (\r{WDW1})
are mutually compatible for the same reason that their classical
counterparts (\r{h}) are. Recall that the worldsheet coordinates
$(\x , \xb )$ appear explicitly only because we have adopted the special
gauge (\r{gaugefix}) identifying the dilaton field with one of the
coordinates. In other words, this
choice of gauge makes the quantum state $\Phi$
time-dependent through the identification of time with
the ``clock field" $\rho$. We note that this long suspected mechanism
for the emergence of time from the ``timeless" WDW equation
here comes almost for free (see e.g. \c{Isham} for a review and further
references). In a covariant treatment the gauge choice (\r{gaugefix})
would have to be undone,
and the full quantum state would be a functional
of $\rho$ rather than a function of the worldsheet coordinates.

\subsection{Hilbert space, physical states and quantum observables}
For obvious reasons we require the total physical Hilbert space $\H$ 
to be unitary\footnote{Although in a Euclidean formulation the
requirement of unitarity should really be replaced by some version
of Osterwalder Schrader reflection positivity.}.
Accordingly, in a given
isomonodromic sector we assign to every $\t_j$ some 
unitary representation space of $SL(2,{\bf R})$ and then take the 
direct product of these spaces. We recall that all $\t_j$ are 
assumed to be purely imaginary; admitting arbitrary complex $\t_j$
would necessitate replacing $SL(2,\R)$ by $SL(2,{\bf C})$, since
for non-self-conjugate pairs with $\t_j=-\bar{\t}_k$ 
($\t_j\neq\t_k$) the matrices $A_j$ are complex. 
As is well known unitary representations of non-compact 
groups are infinite dimensional. For $SL(2,\R)$ 
(or $SU(1,1)$) one distinguishes the continuous 
(principal and supplementary) series and the 
discrete series representations \c{Lang} (see also App. A for a summary),
whereas for $SL(2,\C)$ no discrete unitary representations exist.
In contradistinction
to the compact case, where we would have to deal with the 
standard spin $s$ representations of $SU(2)$ utilized in \c{KN2}
we must decide therefore
which unitary representations of $SL(2,\R)$ to use here.
Although we will be mainly concerned with discrete series representations
in the remainder, we would like to emphasize that our only reason
for ignoring the continuous series representations here is
that the technology for solving KZ equations for them is not yet 
sufficiently developed. As we just explained the necessity of
including such representations becomes already evident when 
one tries to extend the present treatment to arbitrary complex $\t_j$.

The total Hilbert space containing all $N$-soliton sectors
is obviously quite large. As explained above, we can view the
quantized $N$-soliton sector as a system of $N$ compact (for $SU(2)$)
or non-compact (for $SU(1,1)$) ``spins" located at the points 
$(x,\rho) =(w_j,0)$, where the classical solutions of 
(\r{ee}) and  (\r{h}) generically have 
singularities on the worldsheet \c{KM}. Since we expect all
worldsheet points to be equivalent, it seems that we would have to
assume these representations to be the same for all $w_j$.
However, such configurations would not give
all possible classical solutions in the classical limit.
This suggests that we should not assign one particular
representation to every point, but instead the (formal) direct sum
of all representations
\be
\H(w_j):=\bigoplus_{s} \H^{(s)}
\la{Hilbert0} \ee
(of $SU(2)$ or $SU(1,1)$ if $w_j$ is real and $SL(2,{\bf C})$ otherwise).
Then the Hilbert space associated with 
an isomonodromic sector is parametrized by the set $(w_1,...,w_N)$ 
(which is invariant under complex conjugation) 
and given by the direct product
\be
\H^{(N)} \big(\{w_j\}\big) := \bigotimes_{j=1}^N \H(w_j)  
\la{Hilbert1}  \ee
The full Hilbert space $\H$ should contain all these spaces 
as subspaces, i.e. we should demand that
\be
\H^{(N)} \big(\{w_j\} \big) \subset \H    \la{Hilbert2}
\ee
for all possible choices of $N$ and $w_j$. For any two 
disjoint sets of data $\{w_j\}$ and $\{w'_k\}$ with respective
soliton numbers $N$ and $N'$, we can construct a new Hilbert space
describing both configurations by taking the direct product
\[ \H^{(N+N')}\big(w_j,w'_k\}\big) :=
   \H^{(N)}\big(\{w_j\}\big) \bigotimes
   \H^{(N')}\big(\{w'_k\}\big) .   \la{Hilbert3}     \]
If the sets are not disjoint, we simply include the overlapping
factors only once in the product. In this way, we ensure that the various
subspaces are consistently embedded in a partially ordered sequence
of Hilbert spaces, and we can therefore define the
total Hilbert space $\H$ as the inductive limit of the subspaces 
contained in it. Superficially, $\H$ thus looks like a Fock space
with $N$ playing the role of a particle number operator, but
matters are complicated by the fact that there is a continuous
infinity of parameters on which $\H$ may depend, so we cannot 
expect the Hilbert space $\H$ to be separable. This construction
would be reminiscent of similar constructions in the context of
recent work on the loop representation of canonical gravity \c{Lew};
at the classical level it is related to the problem of
whether and in what sense the isomonodromic solutions exhaust 
the full phase space of dimensionally reduced Einstein gravity.
In summary, the structure of the full Hilbert space $\H$ is 
rather complicated and remains to be fully elucidated.

Physical states $\Phi$ must satisfy the quantum analog of
\Ref{WDW}, i.e. the WDW equations
\be \Cx \Phi = \Cxb \Phi = 0 \la{quantumWDW} \ee
By \Ref{C=d/dx} these equations are equivalent to
\be
\f{d\Phi}{d\x}=\f{d\Phi}{d\xb} = 0
\ee
Therefore the physical states are independent of the coordinates;
we thus have a rather simple realization of the idea that
physical states in quantum gravity should be invariant under
the full set of $2d$ coordinate transformations! The usefulness
of this observation relies essentially on the  
fact that the coordinate dependence enters
essentially only via the spectral parameter $\t$ and the constraints
$\Cx$ and $\Cxb$ are represented by highly non-trivial operators. 
Quantum observables $\cO$ by definition commute weakly 
with the constraint operators $\Cx$ and $\Cxb$, i.e.
$$ \big[ \Cx \, , \, \cO \big] \Phi
= \big[ \Cxb \, , \, \cO \big] \Phi = 0 $$
for any physical state $\Phi$. Thus from any such state we can
obtain another physical state by application of the operator $\cO$.
In section 3.5 we have given a large variety of classical observables
based on the monodromy matrices $M_j$.
Since these present no ordering problems of any kind for finite $N$, 
we can straightforwardly take them over to the quantum theory.
In other words, the quantum monodromies are obtained from
\Ref{monodromy} by promoting $A(\t)$ to an operator in accordance 
with \Ref{Aquant}. As before we shall consider only the restricted
set $\sbo$  consisting of (the quantum versions of)
the classical integrals of motion $\tr A_j^2$ and $ A_{\i}$. 
As we will see the quantization of the Schlesinger 
equations in terms of $A_j$ leads to the KZ equations; then the
quantum monodromies turn out to coincide with the monodromies 
of the KZ equations up to similarity transformations and to carry
representations of certain quantum group \c{Drinfeld,Schomerus}.

Since we are working with unitary representations the natural
scalar product on any subspace $\H^{(N)}$ is automatically
positive definite (negative norm states would, however, arise in
a fully covariant treatment, i.e. prior to the conformal
gauge fixing \Ref{gaugefix}, and would have to be eliminated by
a suitable gauge constraint). Furthermore, owing to the 
coordinate independence of the physical states \Ref{quantumWDW} 
this scalar product is invariant under the full diffeomorphism group
when restricted to physical states because for two such states 
$$\f{d}{d\x}\langle \Phi_1|\Phi_2\rangle=
\langle \Cxb \Phi_1|\Phi_2\rangle+ \langle \Phi_1|\Cx\Phi_2\rangle =0$$
In a covariant formulation, 
we would have to integrate over the fields $\x(z)$ and $\xb(\zb)$ 
with a suitable measure, which would presumably include a 
Faddeev Popov determinant to ensure diffeomorphism invariance, 
together with a $\d$-functional for gauge fixing. 
Given observables $\cO_1, \dots , \cO_n \in\sbo$ and any non-trivial
physical state $\Phi$ we can thus compute the correlators
\be
\big \langle \cO_1 \cdots \cO_n \big \rangle =
\f{\langle \Phi | \cO_1 \cdots \cO_n | \Phi \rangle}
 {\langle \Phi | \Phi \rangle} \la{correlator} \ee
By construction, such expectation values are invariant under 2D
diffeomorphisms and therefore meaningful objects
in quantum gravity.

\subsection{Solving the Wheeler-DeWitt equation}

The central task is now to solve the quantum constraints 
\Ref{quantumWDW}, which can be written out as
\be
\Big( \k_{\x} -  \Hx \Big) \Phi = \Big( \k_{\xb} - \Hxb \Big) \Phi = 0.
\la{WDW1}
\ee
where $\Phi$ is the full quantum state. 
Using \Ref{cofa}, equations (\r{WDW1}) take the form
\begin{equation}
-i\hbar\f{\p\Phi}{\p\x} = H^{(\x)} \Phi   \;\;\;\;\;\;\;
-i\hbar\f{\p\Phi}{\p\xb} = H^{(\xb)} \Phi
\label{WDW2}
\end{equation}
where, in the $N$-soliton sector, $\Phi$ is an ${\H}^{(N)}$-valued
function of $(\x , \xb )$. Readers may wonder at this point
why these equations are {\em first order}, since the usual
WDW equation is a second order (functional) differential equation.
This feature is explained by the fact that the first order equations 
(\r{WDW2}) arise due to the separation of the theory into 
left and right moving sectors. This is just a more complicated 
realization of the simple fact that the free wave equation
in two dimensions can be reduced to two first order equations.

To simplify matters we shall not utilize the full Hilbert space
\Ref{Hilbert1} but only consider functionals that live
in the $N$ soliton subspace
$$
\H^{(N)} \big(\{s_j,w_j\}\big) := \bigotimes_{j=1}^N \H_j\;\;\;\;\;,\;\;\;\;
\H_j\equiv \H^{(s_j)}
$$
with fixed $w_j \in {\bf R}$ and $s_j$. For the rest of this and the
following sections we will work entirely with the $SU(1,1)$ 
basis from now, because this permits an easy passage to 
the compact group $SU(2)$. Furthermore, we will restrict
attention to the discrete unitary representations, leaving the
consideration of continuous representations for future work.
Explicit solutions of (\r{WDW2}) on the subspace $\H^{(N)}$
may be obtained by exploiting the close link between (\r{WDW2}) 
and the KZ equations \c{KZ} for $SL(2,{\bf R})$ which read
\be
\f{\p \Phi_{KZ}}{\p\t_j}=-i\hbar\sum_{k\neq j}\f{\Omega_{jk}}
{\t_j-\t_k} \Phi_{KZ}
\la{KZ}\ee
with an ${\H}^{(N)}$-valued function $\Phi_{KZ}(\x,\xb)$. Here 
\be \Omega_{jk}: =\ft12 h_j\otimes h_k + f_j\otimes e_k +
e_j\otimes f_k    \la{Casimir1}   \ee
is a linear operator which for $j\neq k$ acts non-trivially only in
$\H_j$ and $\H_k$ and as the unit operator on the other spaces
(the operators $\Omega_{jk}$ retain their form when we replace
the $SU(1,1)$ generators by the $SL(2,\R)$
generators $\e_j,\fg_j,\h_j$).

Solutions of \Ref{KZ} for $SU(2)$ were apparently first 
constructed in \c{FD,SV}. The
adaptation of these results to the positive discrete series 
representations of $SL(2,\R)$ reviewed in the appendix works as follows:
substituting expressions \Ref{efh} for generators $e_j,f_j,h_j$
in terms of variables $z_j$, we see that $\O_{jk}$ are 
differential operators of the second order, and
the KZ equations \Ref{KZ} can be realized as
the following system of linear partial differential equations 
for the wave function $\Phi(\t_1,...,\t_N ;z_1,...,z_N)$:
\be
\f{\p\Phi}{\p\t_j}= -i\hbar \sum_{k\neq j} \f{1}{\t_j-\t_k}
\left\{-(\tz_j -\tz_k)^2 \f{\p^2\Phi}{\p\tz_j\p\tz_k} 
  +(\tz_k-\tz_j)(s_j\f{\p\Phi}{\p\tz_k}-
   s_k\f{\p\Phi}{\p\tz_j}) + \f{s_j s_k}{2}\Phi
\right\}
\la{KZD}\ee
Solutions of these equations corresponding to the positive discrete
series correspond to functions which for all $z_j$ are 
holomorphic in the upper half plane. They are described by 
\begin{Theorem}
The following expression satisfies equations (\r{KZ}):
\be
\Phi_{KZ}= \oint_{\D_1} dv_1 \cdots 
     \oint_{\D_M} dv_M \, W \,\phi
\la{SolKZ}\ee
where 
\be
\phi := e(v_1)...e(v_M)|p\rangle
\la{phi}\ee
are the ``off-shell" Bethe states introduced in \Ref{Bethe} and 
the function $W$ is defined by
\be
W\big( \{\t_j,v_r \} \big)
:=\prod_{1\leq k < j\leq N} (\t_j-\t_k)^{-i\hbar s_j s_k/2}
\prod_{1\leq s < r\leq M} (v_r-v_s)^{-2 i \hbar}
\prod_{r=1}^{M}\prod_{j=1}^{N}(\t_j-v_r)^{-i \hbar s_j} 
\la{W}\ee
and assumed to be single-valued on the cycle $\Delta$ in $\C^M$
consisting of a family of $v_r$-dependent contours $\Delta_r$ in $\C^M$ 
having empty intersection with the hyperplanes 
$v_s=v_r$ for $r\neq s$ and the $(\x,\xb)$-dependent 
hypersurfaces $v_r = \t_j$, where $r,s=1,...,M$ and $j=1,...,N$.
\la{KZSol}
\end{Theorem}

{\it Proof} (see \c{SV,B}). In the remainder 
we shall use the shorthand notation
\be
\oint_\D dv \equiv \oint_{\D_1} dv_1 \cdots
       \oint_{\D_M} dv_M    \la{shorthand} \ee
for the multiple contour integral in \Ref{SolKZ}. Let us also define
\be
\phi_r := e(v_1)\cdots e(v_{r-1}) e(v_{r+1}) \cdots e(v_M) |p\rangle
           \la{shorthand2}    \ee
so that
\be
\f{\partial \phi_r}{\partial v_r}=0
\la{dphidv} \ee
since the term $e(v_r)$ is omitted from $\phi_r$.   
Then by a lengthy but straightforward calculation, 
taking into account the definiting conditions \Ref{Bethe1},
one shows that 
\be
H_j\phi =(i\hbar)^2
 \Big(\alpha_j \phi -\sum_r \f{\beta_r}{\t_j-v_r}e_j \phi_r \Big)  
\la{Hjp}\ee
where
\be
H_j=(i\hbar)^2\sum_{k\neq j} \f{\O_{jk}}{\t_j-\t_k} 
\la{HjKZ}\ee
are the KZ Hamiltonians entering (\r{KZ}). Here we have defined the functions
\be
\alpha_j= \sum_{i\neq j}\f{s_i s_j}{2(\t_j-\t_i)} -\sum_r\f{s_j}{v_r-\t_j}
\la{al}\ee
\be
\beta_r=\sum_{s\neq r}\f{2}{v_r-v_s}+\sum_j \f{s_j}{v_r-\t_j}
\la{be}\ee
By construction, $W$ in (\r{W}) satisfies
\be
\f{\partial W}{\partial\t_j}=-i\hbar\alpha_j W\;\;\;\;\;\;\;\;\;
\f{\partial W}{\partial v_r}=-i\hbar\beta_r W
\la{Wd}\ee 
Invoking (\r{Hjp}) and (\r{Wd}), we get
$$ \f{\p\Phi_{KZ}}{\p\t_j}+ \f{1}{i\hbar}H_j\Phi_{KZ}
 =\oint_{\D} dv \, \left(-\sum_{r}
\f{\p}{\p v_r}\left\{\f{e_j}{v_r-\t_j}\right\}\phi_r -i\hbar\sum_r
\f{\beta_r e_j}{\t_j-v_r}\phi_r\right)W  $$
$$=-e_j\oint_{\D}dv \, \sum_r \f{\p}{\p v_r}\left\{\f{W}{v_r-\t_j}\phi_r
       \right\} =0 $$
because of \Ref{dphidv}, and there are no boundary contributions
as $\D$ is closed.$\Box$

For a discussion of the completeness of the solutions (\r{SolKZ})
in the case of $SU(2)$ see \c{SV,Var}. 
The cycles $\Delta$, on which    
the solutions depend in addition to the parameters characterizing 
the $N$-soliton Hilbert spaces are generically rather complicated,
especially in the limit $\hbar \rightarrow 0$. For $SL(2,\R)$, 
the space of solutions is infinite dimensional even for fixed 
soliton number $N$ due to the fact that the underlying Hilbert 
space $\H^{(N)}$ is also infinite dimensional:
unlike for $SU(2)$, the basic formula
(\r{SolKZ}) yields non-trivial solutions for arbitrary $M$.

With the solution \Ref{SolKZ} of the KZ equation at hand,
we can now proceed to the construction of the full
WDW functional $\Phi$ solving \Ref{WDW2}.
 
\begin{Theorem}
Let $\Phi_{KZ} \in \H$ be any solution \Ref{SolKZ} of the 
KZ equations (\r{KZ}). Then the $\H$-valued function $\Phi (\x,\xb)$ 
defined by
\be
\Phi = (\x-\xb)^{-\ft14 \hbar^2\si (\si -2)}
\prod_{j=1}^{N} \bigg( \f{\partial \t_j}{\partial w_j}
      \bigg)^{-\f{1}{4}\hbar^2 s_j (s_j-2)} 
 \Phi_{KZ} 
   \Big( \big\{ \t_j \big\} \Big)  
\la{Sol1}\ee
with
\be \si := 2M + \sum_j s_j    \la{sinfinity}  \ee
satisfies the WDW equation (\r{WDW2}).
\la{link1}
\end{Theorem}
{\it Proof.} From (\r{Aquant}) we find 
$$ \tr (A_j A_k) =-\hbar^2 \Omega_{jk} $$
Moreover, one can check the following relation between the total
Hamiltonians $\Hx$, $\Hxb$ and the KZ-Hamiltonians $H_j$  \Ref{Hj} 
which resembles the classical relation \Ref{calc1} between $\Hx$, $\Hxb$ and 
the Schlesinger Hamiltonians. For instance,
\be
 \Hx=\sum_j H_j\t_{j\x}-\f{1}{\x-\xb}\sum_{j} \tr(A_j^2)\left(\f{1}{2}-
\f{1}{(1-\t_j)^2}\right) +\f{1}{2(\x-\xb)}\tr\Big(\sum_j A_j\Big)^2 
\la{proof}\ee
The second term on the r.h.s. of  \Ref{proof} involves  
the Casimir operators of the respective $SL(2,\R)$ representations
and acts on the off-shell Bethe states according to  
\be
\tr (A_j^2)\, \phi=  (i\hbar)^2 \O_{jj} \, \phi
   = - \ft12 \hbar^2s_j(s_j-2) \phi
\la{Cas}\ee
Writing
$$ -A_\i \equiv \f{i\hbar}{2} \pmatrix{\h_\i & 2\e_\i \cr 2\fg_\i & -\h_\i \cr}
= \f{i\hbar}{4}
\pmatrix{ 1 & -i \cr -i & 1}
\pmatrix{ h_\i & 2e_\i\cr 2f_\i & -h_\i \cr}
\pmatrix{ 1 & i \cr i & 1}
 $$
we have
\be h_\i \phi \equiv \sum_{j=1}^N h_j\phi =
  (2M + \sum_j s_j ) \phi \equiv \si \phi  \la{hinfinity}  \ee
For the last term of \Ref{proof} we then get
\be
\tr\Big(\sum_j A_j\Big)^2 \phi = (i\hbar)^2\sum_{j,k}\O_{jk}\phi=
 - \hbar^2\Big(\ft12 \si (\si-2) \phi +
 \sum_{j,r}\beta_r e_j\phi_r  \Big) 
\la{Castot}\ee
where the functions $\beta_r$ and the states $\phi_r$
are defined in \Ref{be} and \Ref{shorthand2}, 
respectively, and the off-diagonal last term on the r.h.s is
due to the action of the operator $e_j f_j$.
The r.h.s. of \Ref{Cas} contributes in an
obvious way to \Ref{Sol1}. As for \Ref{Castot}, the second term
on the r.h.s., which is not diagonal, does not contribute to
$\Hx \Phi$ because by eq. \Ref{Wd} and by partial integration 
(with $\partial \Delta = \emptyset$), it gives a term
inside the integral \Ref{SolKZ} proportional to 
$$  \oint_{\D} dv
            \sum_{j,r} \f{\p W}{\p v_r} e_j \phi_r 
    = - \oint_{\D} dv \sum_j e_j W \f{\p \phi_r}{\p v_r}  $$
which vanishes by \Ref{dphidv}. In this way we arrive at
\be
\tr \Ai^2 \Phi_{KZ}= - \ft12 \hbar^2 \si (\si -2) \Phi_{KZ} \la{AiPhiKZ}
\ee
which is diagonal; therefore all terms depending explicitly on
$(\x,\xb)$ can be integrated straightforwardly.  $\Box$

We note the strong similarity of \Ref{Sol1} with
the classical formula \Ref{link3}: the prefactor is essentially 
the same except that we have replaced the classical expressions
in the exponent by the eigenvalues of the corresponding operators. 
The factor involving $(\x -\xb)$ can be absorbed into a renormalization 
of $\Phi$ by appropriate choice of the function $f(\x,\xb)$ 
in \Ref{cofa}. The state $\Phi_{KZ}$ can therefore be regarded as the 
quantum analog of the classical $\tau$-function\footnote{Alternatively, 
one could identify the quantum $\tau$-function with the
corresponding evolution operator which is a "matrix" whose columns 
constitute an orthonormal basis of states $\Phi_{KZ}$ in the space 
of solutions of KZ equations.}.

Formula \Ref{Sol1} gives the general solution of \Ref{WDW2} for 
both $SU(1,1)$ (and therefore $SL(2,\R)$) and $SU(2)$ for the respective
lowest weights $s_j$. For $SU(2)$ the theorem was already stated
without proof in \c{KN2} whereas the non-compact case was not
considered there. This leaves us with the task of
translating the classical coset conditions of section 3.3 
into the quantum theory. However, we cannot directly generalize
the condition $g(\x,\xb)\in \coset$ because we do not know the 
proper definition of the quantum operator corresponding to $g(\x,\xb)$.
This is why in section 3.3 we explored different ways to 
formulate the coset constraint. The similarity between
$\tau$ and $\Phi_{KZ}$ suggests the necessary conditions 
\be
\Big(\sum_{j=1}^N A_j \Big) \Phi_{KZ}=0 
\la{qAi}\ee
and
\be
\Phi_{KZ}\bigg(\f{1}{\t_1},...,\f{1}{\t_N}\bigg)=\Phi_{KZ}(\t_1,...,\t_N)
\la{tau1}\ee
These imply the following restrictions on the parameters of $\Phi_{KZ}$, 
and hence on the full WDW wave functional $\Phi$ by Thm.~\r{link1}.
\begin{Theorem}
The solution $\Phi_{KZ}$ in \Ref{SolKZ} of the KZ equations
satisfies the symmetry conditions \Ref{qAi} and (\r{tau1}) if $N=2n$,  
$\t_{j+n}={\t_j}^{-1}$ and $s_{j+n}=s_j$ ($j=1,...,n$), 
\be
s_\i\equiv 2M + \sum_{j=1}^{N} s_j =0
\la{coset1}\ee
and the cycle $\Delta$ in (\r{SolKZ}) is invariant with respect to the
continuous deformation of $\t_j$ into ${\t_j}^{-1}$.
\la{symm}
\end{Theorem}

Unfortunately, there is a simple argument showing that \Ref{coset1} 
cannot be satisfied for the non-compact theory with the
discrete representations of $SU(1,1)$ or $SL(2,\R)$, unlike
for the compact group $SU(2)$ (for which $s_j\leq 0$).
Namely, for the non-compact groups
we have $s_j > 0$ and condition \Ref{coset1} can never be met;
switching to the negative discrete series, for which $s_j <0$,
does not help because $M$ becomes negative due to the interchange
of $e_j$ and $f_j$. Therefore it appears that
the discrete representations of $SU(1,1)$ are unsuitable
for the task at hand, at least as long as we do not make
simultaneous use of the positive {\em and} negative series.
This conclusion is confirmed by an analysis of the sign of 
the Casimir operator (cf. next section) which shows explicitly
that the known classical solutions are associated with the principal 
series representations of $SL(2,\R)$ or $SL(2,\C)$. 
Moreover, let us repeat that for $\t_j=-\overline{\t_k}$, 
when $\H_j$ carries a unitary representation of $SL(2,\C)$, 
the discrete series are altogether absent, so the consideration
of the principal series cannot be avoided in any case.
Finally, a complete treatment of the quantum coset constraints
will require the implementation of \Ref{step5} at the quantum level.

As an aside we observe that the ``ultra quantum limit"
$\hbar\ra\i$ corresponds to the semi-classical limit
of the WZNW model (and vice versa). Heuristically,
this limit is dominated by the stationary points of the integral
which are determined by the equations $\beta_j=0$, or, equivalently,
\be
\sum_{s\neq r}\f{2}{v_r-v_s} +\sum_j\f{s_j}{v_r-\t_j}=0
\la{Betheeq}   \ee
These are nothing but the so-called Bethe equations \c{FT, Korep}
which diagonalize the ``matter Hamiltonians" $H_j$ on the 
Bethe states \Ref{Bethe}, except that the Bethe parameters $v_r$ 
here depend on the coordinates because $\t_j = \t_j(\x,\xb)$.
Consequently, in the limit $\hbar\ra\i$, the quantum states become
more and more sharply peaked about the ``on-shell" Bethe states.

\subsection{Remarks on the classical limit}

Although we have not been able to solve the coset constraints
and to find the associated constrained quantum states,
we can still discuss some general features of the corresponding
classical solutions. The solutions based on the discrete 
representations possess some rather unusual features from 
this point of view. To see this let us compute
the possible eigenvalues of the operator $\tr A_j^2$, whose 
classical counterpart is positive for all known classical solutions 
satisfying the conditions of Thm.~\ref{main}, being equal to
$\tr \, T_j^2$ by \Ref{Aj}. In fact, no classical solutions
with $\tr A_j^2 < 0$ are known, although we are aware of no reason of
principle forbidding their existence. In the quantum theory we get
\be
\tr A_j^2 = -\hbar^2\Big(\ft12 h_j^2 + e_j f_j +f_j e_j \Big)
\la{Casim} \ee
which is equal to  $-\f{1}{2}\hbar^2 s(s-2)\leq 0$. 
The operator in parentheses is just the Casimir operator which 
is positive definite for $SU(2)$ (since $f_j = e_j^{\dagger}$), 
but in general not for $SU(1,1)$, although the eigenvalues of \Ref{Casim} 
are still positive for discrete representations.

To study the classical limit of the quantum states we have 
to consider the expectation values of the various observables
and their $\hbar \ra 0$ limits. Although it is a difficult
problem to do this in full generality, it is at least clear
that in order to end up 
with a non-trivial classical solution the expectation values 
$\langle \tr A_j^2\rangle$ must stay finite in the limit 
$\hbar\rightarrow 0$. This can be achieved by letting
$\hbar\rightarrow 0$ and $s_j\rightarrow \i$ 
in such a fashion that the product $\hbar s_j$ remains finite, i.e.
\be
\hbar s_j \rightarrow s_j^{(0)}\neq 0 \;\;\;\;\;{\rm as}\;\;\;\;\;
\hbar\rightarrow 0
\la{scl}\ee
Then
\be
\lim_{\hbar \ra 0} \tr A_j^2 = -\ft12 (s_j^{(0)})^2 < 0
\la{scl1} \ee
As we just pointed out $\tr A_j^2$ is positive for all 
known classical solutions. Moreover, multisoliton and 
algebro-geometrical solutions \cite{KOR} require non-self-conjugate pairs 
$w_j=\bar{w}_k$, whereas solitons ``sitting" on the real axis 
$w_j\in \R$ (i.e. $\t_j\in i\R$) are absent. This fact is directly 
related to the absence of representations of $su(2)$ 
with negative Casimirs eigenvalue. By contrast,
such representations do exist for $SL(2,\R)$ and $SL(2,\C)$.
In the limit \Ref{scl}, the quantum state \Ref{Sol1} becomes  
\be
\Phi_{KZ} \sim \prod_{1 \leq k <j \leq N}
 \big(\t_j - \t_k \big)^{i s_j^{(0)} s_k^{(0)}/2\hbar} 
 \oint_{\Delta} dv \, \prod_{r=1}^M \prod_{J=1}^N
 \big( \t_j - v_r \big)^{i s_j^{(0)}} \,
 |p;v_1,\dots , v_M \rangle 
\la{semiclassical} \ee
If this limit could be properly controlled one would have a better 
handle on the ultraviolet divergences that would appear in a conventional
perturbative treatment based on an expansion about $\hbar =0$,
but are ``invisible" in the isomonodromic sectors.

While the standard techniques to solve the KZ equations for Verma modules 
did yield solutions of the WDW equation for the compact theory,
they do not shed much light on the physically more interesting
situation corresponding to lowest or highest weight representations 
of $SL(2,\R)$ because we are unable to satisfy 
the coset constraints \Ref{coset1} with discrete representations.
Unlike for $SU(2)/U(1)$ we know that multisoliton 
solutions with real $w_j$ do exist for $\coset$; one example
is the Kerr-NUT solution, whose existence is related to the
existence of continuous representations
of $su(1,1)$ with negative Casimir.
For conjugate pairs $w_j=\bar{w}_k$ the associated ``spins" $A_j$
and $A_k$ again give conjugate representations of $sl(2,\C)$, which
give rise to the Kerr-NUT solutions with naked singularities. 

Let us sketch how one might go about constructing a ``realistic" quantum 
state corresponding to a ``quantum Kerr black hole": select
four singular points $\big\{ \t_1 \equiv \t (w_1)\, , \,
\t_2 \equiv \t (w_2)\, ,\, \t_3 = \t_1^{-1} \, ,\, \t_4 = \t_2^{-1}\big\}$
with $w_1=-w_2 \in \R$ and assume the spaces
$\H_j$ to be the same representation from the principal 
series with Casimir eigenvalue equal to $-\f{1}{2}(1+q^2)$
(here $s=1+iq$) for $j=1,2,3,4$. 
Substituting the differential operator representation 
of Appendix A for $e_j, f_j, h_j$
(which is valid for all unitary representations of $SU(1,1)$)
into \Ref{KZ} and \Ref{Casimir1} we obtain the following
partial differential equations for the state
$\Phi(\t_1,...,\t_4;\tz_1,...,\tz_4)$ 
\be
\f{\p\Phi}{\p\x}= -i\hbar \sum_{k \neq j} \f{1}{(1-\t_j)(1-\t_k)}
\left\{-(\tz_j -\tz_k)^2 \f{\p^2\Phi}{\p\tz_j\tz_k} 
+s(\tz_k-\tz_j)\big(\f{\p\Phi}{\p\tz_k}-
\f{\p\Phi}{\p\tz_j}\big) + \f{s^2}{2}\Phi \right\}
\la{WDWG}\ee
The wave functionals 
\Ref{SolKZ} above correspond to solutions of this equation
with functions which are holomorphic in the upper half planes 
${\rm Im}\, \tz_j \geq 0$ and normalizable with respect to the 
measure \Ref{measuredis} in  all the  variables $\tz_j$ simultaneously
(however, these solutions do not satisfy the coset constraints). 

For continuous representations, the variables 
$\tz_j$ are real and again no solutions of \Ref{WDWG} are known. 
Together with the constraint \Ref{qAi}, the system
\Ref{WDWG} reduces to a single partial differential equation which
originally arose  in Liouville theory \c{Liouville} where it 
governs the four-point correlation function.
Explicit solutions of this equation are not known so far,
but should certainly exist.

\section{Stringy aspects}
\setcounter{equation}{0}
At this point it is fitting to emphasize
the intriguing analogies of our results with certain aspects
of string theory and the tantalizing hints of a new type
of theory which could emerge upon a ``stringy" 
reinterpretation of (a supersymmetric version of)
dimensionally reduced quantum gravity along the lines 
already suggested in \c{Nic87}. 
The similarity of the dilaton field $\rho$ 
and the logarithm of the conformal factor 
with the longitudinal target space coordinates $X^+$ and $X^-$, 
respectively, was already pointed out in \c{Nic}. Furthermore,  
the choice of Weyl canonical coordinates in \Ref{gaugefix}
corresponds to the lightcone gauge fixing condition $X^+ = \tau$ in 
string theory; in \c{KN3} it was shown that this analogy remains valid for
higher genus worldsheets if one identifies $\rho$ with the globally 
defined (lightcone) time coordinate introduced by Mandelstam 
to describe string scattering in the lightcone gauge. The fact that through 
this choice of gauge the field variables $\k_{\x}$ and $\k_{\xb}$
become canonically conjugate to the worldsheet coordinates
is also in accord with this interpretation. 
The transmutation of the conformal factor into a longitudinal 
target space degree of freedom has also been proposed 
in Liouville theory (i.e. subcritical string theory) \c{Pol}; 
however, the Lagrangians considered for this purpose apparently
do not arise from dimensional reduction of Einstein's theory.

We have also remarked that the WDW equations \Ref{WDW1} should
be considered on a par with the Virasoro constraints of string
theory as they state nothing but the vanishing of the off-diagonal
components $T_{\x \x}$ and $T_{\xb \xb}$ of the full energy
momentum tensor on the physical states, where the full energy momentum
tensor is defined to include the gravitational contribution
$\p_\x k \equiv 2i \p_\x k \p_\x \rho$ (or its hermitean conjugate),
again in analogy with Liouville theory. 
Thus instead of regarding the expressions \Ref{Sol1}
as solutions of $2d$ matter coupled quantum gravity,
we could alternatively interpret them as physical states
in some higher dimensional target space. 
The operators $e(\t)$ and $f(\t)$ would play the 
role of transverse creation and annihilation
operators, respectively. We note that the relevant ``oscillators"
are not the Fourier coefficients of the string target 
space coordinates with respect to the left and right moving
worldsheet coordinates $\x$ and $\xb$, but rather based
on an expansion with respect to the spectral parameter $\t$
which appears to be the truly fundamental variable.
Our hypothetical string would thus be neither closed nor
open, but ``unidexterous". The similarity  
between the contour integrals over products of ``oscillators" 
appearing in \Ref{Sol1} and the corresponding expressions
giving physical string states in terms of DDF operators 
is noteworthy (in fact, our expressions are almost identical
with the integrals for correlators of the $c<1$ minimal 
conformal models \c{FD}). 

Of course, we would not expect the new ``string" to be automatically
consistent; instead consistent models should satisfy further
constraints such as absence of anomalies (which is the criterion 
singling out the critical string theories). As is well known,
the crucial consistency test in the lightcone gauge
is the closure of the Lorentz algebra (see e.g. \c{GSW}),
and in addition to defining the target space we will have  
to look for the analog of the Lorentz algebra. The obvious candidate
for the transverse subgroup for the model investigated here
is the $SO(2)$ subgroup of the Ehlers group $SL(2,\R)$ 
generated by $\Ai$ (see \Ref{sum}), but the extension to a 
full Lorentz group remains to be found.

One of the outstanding problems of midi-superspace
quantum gravity is to find a symplectic (canonical) realization 
of the Geroch group \c{Ger}. Classically this is simply the 
group of ``dressing transformations" which add extra regular 
singularities to $\Psi(\t)$ with special monodromy data. 
Here we have investigated the isomonodromic sectors separately,
where only representations of $SL(2,\R)$ appear, but the 
general case should involve the affine extension $\widehat{SL(2,\R)}$. 
The appearance of the involution
$\t \ra {\t}^{-1}$ in \Ref{tau0} and \Ref{tau1} moreover suggests that
in a fully covariant formulation the physical states should 
be invariant under the ``maximal compact" subgroup $SO(2)^\infty$
already encountered in \c{Jul,BM,Nic}. While there are 
indications from flat space models that such transformations cannot be 
realized within the conventional canonical approach \c{PL}, 
the situation may be different for our new formulation.
The quantum Geroch group would relate the quantum states
to one another in the same way that the ordinary Geroch group
relates classical solutions; it would mix the isomonodromic sectors
and change the soliton number $N$. The monodromy operators \Ref{mon}  
are the natural candidates for conserved quantum non-local charges.
The work of \c{Drinfeld} on the quantum group structure 
of the monodromy algebra of the KZ equations suggests
that the Geroch group could become a true quantum group in the 
technical sense of the word (see also \c{Schomerus}).
In the string context the quantum Geroch group would be interpreted as 
a spectrum generating symmetry with the physical states 
belonging to unitary representations of the
relevant non-compact (quantum) group. 

The introduction of interactions between physical string states 
would amount to a ``third quantization" from the $2d$ worldsheet
point of view, such that scattering processes involving different 
physical states would correspond to the interaction of 
different $2d$ ``universes". However, just as the construction of a
proper string field theory requires more than the Virasoro constraints,
such an interpretation of our model would involve essentially
new elements beyond the WDW equations \Ref{WDW1}. In particular,
it would necessitate repeating the analysis of this paper for
higher genus worldsheets, building on earlier results of \c{KN3}.
Readers may appreciate the resemblance of these ideas
with recent attempts to understand the possible loss of quantum
coherence and the emergence of baby universes in quantum gravity 
on the basis of certain $2d$ models \c{Rub};
however, our intentions here really go in the opposite direction 
as we wish to build a new kind of string theory from these models
rather than to treat string theory as an ancillary model
to understand features of $2d$ quantum gravity. 

Finally, the manifest split into left and right moving sectors 
put in evidence by the automatic compatibility of \Ref{1} and the 
mutual commutativity of the Hamiltonians \Ref{Ham} 
as well as the canonical constraints \Ref{constraint1} is strongly
reminiscent of holomorphic factorization in string theory. 
It suggests that our formulation is the natural
starting point for studying the reduction to one dimension.
In analogy with string theory, which can be regarded as a $2d$
field theory composed of two one-dimensional (chiral) halves,
such a reduction would not really be a dimensional reduction 
to one dimension. One might even argue that the $2d$ theory
{\it is} already ``one-dimensional" in that the 
$(\x,\xb)$-dependence essentially enters only via the
(analytic) dependence on one complex variable $\t$.
In contrast to a naive dimensional reduction of the original
theory, which would just leave us with trivial plane waves,
the rich structure of stationary axisymmetric or colliding plane wave 
quantum gravity would be entirely preserved in this scheme.
This also indicates that the spectrum of the new theory would 
contain many more excitations than the ordinary string because
the structure of unitary representations of $SL(2,\R)$ and 
other non-compact groups (see \c{murat}) is considerably 
more intricate than the ``linear" harmonic oscillator spectra 
of string theory, leaving room for myriad solitonic excitations 
at the quantum level.

{\bf Acknowledgements.} We would like to thank A.~Ashtekar, H.~Babujan,
M.~G\"unaydin, A.~Kitaev, J.~Lewandowski, H.-J.~Matschull,
A.~Mironov, A.~Morosov, H.~Samtleben, V.~Schomerus, 
M.~Semenov-Tian-Shansky, F.~Smirnov and J.~Teschner,
who have at various stages contributed insightful comments
to the present work. D.~K. acknowledges the support of the
A.v.~Humboldt Stiftung and Deutsche Forschungsgemeinschaft
under contract No. Ni 290/5-1.

\begin{appendix}

\section{Unitary representations of $SL(2,{\bf R})$}
For the convenience of the reader we here summarize some pertinent 
results about unitary representations of  $SL(2,\R)$;
see \c{Lang} for details and further information.
All representation spaces $\H^{(s)}$ can be realized as Hilbert spaces of 
functions $\phi (z)$ of one (complex or real) variable $z$.
The action of the group element $G\in SL(2,{\bf R})$
\be
   G= \pmatrix{a&b \cr c & d \cr}
\ee
(where $ad- bc =1$) on any such function is given by
\be
  T_s(G) \phi(\tz)=\phi\left(\f{a\tz+b}{c \tz+ d }\right)
    (c\tz+ d)^{-s} 
\ee
if $s$ is integer; otherwise (i.e. for continuous representations)
we have $|c\tz+d|^{-s}$ instead of  $(c\tz+d)^{-s}$
on the r.h.s. of this formula.
For all representations, the Chevalley generators
with commutation relations
$$ [\h,\e]=2\e\;\;\;\;\;\;\; [\h,\fg]=-2\fg\;\;\;\;\;\;\; [\e,\fg]= \h $$
are represented on $\H^{(s)}$ by the differential operators
\be
 T_s (\e)=  \tz^2 \f{d}{d\tz}+ s\tz\;\;\;, \;\;\;
T_s (\fg) =-\f{d}{d\tz}\;\;\;, \;\;\;
T_s (\h)=2\tz\f{d}{d\tz}+s
\la{gener}\ee
where the parameter $s$ must satisfy the constraints given below.
These operators are antihermitean with respect to the
scalar products given below. 
The Casimir operators $ T_s \big(\f{1}{2} \h^2+ \e\fg +\fg\e\big) $
is always diagonal; by direct computation
one easily verifies that its eigenvalue is $\f{1}{2}s(s-2)$.

All unitary irreducible representations of $SL(2,{\bf R})$ 
(with exeption of so-called limit of the discrete series corresponding to 
$s=1$) are contained in the following list.

\begin{itemize}
 
\item
For the principal series the functions $\phi (z)$ live on the real line, i.e.
$\tz\in {\bf R}$, and the scalar product is the ordinary $L^2 (\bf R)$
product (which is independent of $s$)
\be
(\psi , \phi)= \int_{{\bf R}} 
\overline{\psi (\tz)} \phi (z) d\tz  
\la{measure}\ee
The allowed values for $s$ are $s=1+iq\;,\;\;q\in {\bf R}$,
and the spectrum of the operators $T_s [\e], T_s [\fg]$ and $T_s [\h]$ 
is continuous for all such $s$.

\item
For the supplementary series the functions $\phi(z)$ are again
defined on the real axis, but the scalar product now depends on
$s$ and is given by
$$ (\psi , \phi)_s =\int_{{\bf R^2}} \overline{\psi (z_1)}
 \phi (\tz_2)|\tz_1-\tz_2|^{s-2} d\tz_1 d\tz_2  $$

with $s\in {\bf R}$ and $ 0 < s <2$ (the latter restriction
follows from requiring $(\phi , \phi )_s$ to be positive
for non-zero $\phi$). The spectrum of the operators 
$T_s [\e],\; T_s [\fg],\; T_s [\h]$ is again continuous.

\item
The positive discrete series representations consist of 
the functions holomorphic in the upper half plane
normalizable with respect to the scalar product
\be
 (\psi , \phi ):=  \int_{{\rm Im} z > 0} \overline{\psi(z)}
 \phi (\tz)\big| {\rm Im} \, \tz \big|^{s-2} d\tz d\bar{\tz} 
\la{measuredis}\ee
where we would have to integrate over the lower half plane for the 
the negative series. In order
to ensure single-valuedness of the functions $\phi (\tz)$, 
only discrete values of $s$ are admitted; furthermore, convergence
of the integral at ${\rm Im}\, z =0$ requires $s\geq 2$, so 
we have $s=2,3,\dots$. 
\end{itemize}
To construct an explicit basis of functions for the discrete series
the $SU(1,1)$ basis is more useful; it is realized by the operators
\begin{eqnarray}
T_s(e)&=&\ft12 T_s \big(-i \h + \e+\fg \big)=
\ft12(\tz-i)^2\f{d}{d\tz}+\ft12 s (\tz-i) \nonumber \\
  T_s (f)&=&\ft12 T_s \big( i\h +\e+\fg \big)=
\ft12 (\tz+i)^2\f{d}{d\tz} +\ft12 s (z+i) \la{efg} \nonumber \\
T_s (h)&=& T_s \big(i(\fg-\e )\big)=-i(\tz-i)(z+i)\f{d}{d\tz} -is\tz
\la{efh} 
\end{eqnarray}
We can explicitly check that
$$
\big[ T_s(h)\, ,\, T_s (e)\big]=2 T_s(e)\;\;\;, \;\;\;
\big[T_s(h) \, ,\, T_s(f)\big]=-2 T_s(f)\;\;\;, \;\;\;
\big[ T_s(e)\, ,\, T_s(f) \big]= T_s(h)
$$
as well as
$$T_s(h)=T_s(h)^{\dagger}\;\;\;\;\; T_s(e)=-T_s(f)^{\dagger} $$
with respect to the scalar product \Ref{measuredis}.
The representation space $\H_+^{(s)}$ of the positive discrete series
is spanned by the following functions holomorphic in the upper half-plane:
\be
\phi_s^{s+k}(z)=(\tz-i)^{k} (\tz+i)^{-s-k}\;\;\;\;\;k=0,1,2,....
\la{basisplus}\ee
The generators $T_s (e) ,T_s (f)$ and $T_s (h)$ act 
on these functions as follows:
\be
T_s (h)\phi_s^{s+k} = (s+2k)\phi_s^{s+k}\;\;\; , \;\;\;
T_s (e)\phi_s^{s+k} = i(s+k)\phi_s^{s+k+1}\;\;\; , \;\;\;
T_s (f)\phi_s^{s+k} = ik\phi_s^{s+k-1}
\la{generation}\ee
The lowest weight state corresponds to the function $\phi_s^s$
which is annihilated by $T_s(f)$ and from which all other
functions can be generated by repreated application of the
raising operator $T_s(e)$; also the operator $T_s(h)$ is diagonal
in this basis with eigenvalues $s,s+2, s+4, \dots$.

\end{appendix}

\end{document}